\documentclass[acmsmall]{acmart}
\usepackage[utf8]{inputenc}
\usepackage[T1]{fontenc}
\usepackage{microtype}
\usepackage{graphicx} 
\usepackage{subcaption}
\usepackage{xspace}
\usepackage{simplebnf}
\usepackage{stmaryrd}
\usepackage{fancyvrb}
\usepackage{amsmath}
\usepackage{array}
\usepackage{booktabs}
\usepackage{float}
\usepackage{cleveref}
\usepackage{mathpartir}
\usepackage{mathtools}
\usepackage{varwidth}
\usepackage{tikz}
\usepackage{tabularx}
\usepackage{natbib}
\usepackage{siunitx}
\usepackage[rightcaption]{sidecap}
\sidecaptionvpos{figure}{c} 
\usetikzlibrary{arrows.meta,positioning,calc,automata,quotes}

\graphicspath{ {./images/} }

\newcommand{\fillin}[1]{{\color{black} #1}}

\newcommand{\blue}[1]{{\color[rgb]{0.298,0.471,0.659} #1}}
\newcommand{\orange}[1]{{\color[rgb]{0.961,0.522,0.094} #1}}

\DeclarePairedDelimiter{\denote}{\llbracket}{\rrbracket}
\newcommand{\muskia}{$\mu$\textsf{Skia}\xspace}
\newcommand{\layer}{\textsf{Layer}\xspace}

\newcommand{\srcover}{\textsf{SrcOver}\xspace}
\newcommand{\dstin}{\textsf{DstIn}\xspace}

\newcommand{\skcanvas}{\texttt{SkCanvas}\xspace}
\newcommand{\sksavelayer}{\texttt{saveLayer}\xspace}
\newcommand{\skrestore}{\texttt{restore}\xspace}
\newcommand{\sksave}{\texttt{save}\xspace}
\newcommand{\skdrawrect}{\texttt{drawRect}\xspace}
\newcommand{\skpaint}{\texttt{SkPaint}\xspace}
\newcommand{\sksrcover}{\texttt{SrcOver}\xspace}
\newcommand{\skmultiply}{\texttt{Multiply}\xspace}
\newcommand{\skdstin}{\texttt{DstIn}\xspace}
\newcommand{\skcliprect}{\texttt{clipRect}\xspace}

\newcommand{\SaveLayer}{\textsf{SaveLayer}\xspace}
\newcommand{\Restore}{\textsf{Restore}\xspace}
\newcommand{\Save}{\textsf{Save}\xspace}
\newcommand{\Paint}{\textsf{Paint}\xspace}

\newcommand{\Image}{\textsf{Image}\xspace}
\newcommand{\Shape}{\textsf{Shape}\xspace}
\newcommand{\Color}{\textsf{Color}\xspace}
\newcommand{\Point}{\textsf{Point}\xspace}
\newcommand{\Blend}{\textsf{Blend}\xspace}
\newcommand{\Filter}{\textsf{Filter}\xspace}

\newcommand{\Layer}{\textsf{Layer}\xspace}
\newcommand{\Empty}{\textsf{Empty}\xspace}
\newcommand{\DrawShape}{\textsf{DrawShape}\xspace}
\newcommand{\BlendLayer}{\textsf{BlendLayer}\xspace}
\newcommand{\Draw}{\textsf{Draw}\xspace}
\newcommand{\Clip}{\textsf{Clip}\xspace}
\newcommand{\Transform}{\textsf{Transform}\xspace}
\newcommand{\SrcOver}{\textsf{SrcOver}\xspace}
\newcommand{\DstIn}{\textsf{DstIn}\xspace}
\newcommand{\Luma}{\textsf{Luma}\xspace}
\newcommand{\Transparent}{\textsf{Transparent}\xspace}
\newcommand{\List}{\textsf{List}\xspace}
\newcommand{\Id}{\textsf{Id}\xspace}

\newcolumntype{L}{>{\sffamily}l}
\newcolumntype{M}{>{$}l<{$}}
\newcolumntype{T}{>{\ttfamily}l}

\newcommand{\tvCompilePassN}{\fillin{41}\xspace}
\newcommand{\tvPassN}{\fillin{38}\xspace}
\newcommand{\tvGrindN}{\fillin{3}\xspace}

\newcommand{\graphiteAppleGeoSpeedup}{\fillin{1.12×}\xspace}
\newcommand{\graphiteAppleMaxOptTime}{\fillin{\SI{0.03}{ms}}\xspace}
\newcommand{\graphiteAppleMaxSpeedup}{\fillin{3.04×}\xspace}
\newcommand{\graphiteAppleMaxTotalSpeedup}{\fillin{3.01×}\xspace}
\newcommand{\graphiteAppleNumBench}{\fillin{139}\xspace}

\newcommand{\graphiteAppleNumWithPixDiff}{\fillin{25}\xspace}

\newcommand{\graphiteAppleMaxSpeedupMatched}{\fillin{3.04×}\xspace}

\newcommand{\ganeshAppleGeoSpeedup}{\fillin{1.08×}\xspace}

\newcommand{\nextGraphiteAppleMaxSpeedup}{\fillin{3.31×}\xspace}

\newcommand{\nextGraphiteAppleNumBench}{\fillin{144}\xspace}

\newcommand{\graphiteIntelGeoSpeedup}{\fillin{1.14×}\xspace}

\newcommand{\graphiteIntelNumBench}{\fillin{143}\xspace}

\newcommand{\graphiteIntelMaxSpeedupMatched}{\fillin{3.13×}\xspace}

\newcommand{\ganeshIntelGeoSpeedup}{\fillin{1.07×}\xspace}

\newcommand{\ganeshIntelMaxSpeedupMatched}{\fillin{1.92×}\xspace}

\newcommand{\N}{\fillin{100}\xspace}

\setcopyright{cc}
\setcctype{by}
\acmDOI{10.1145/3839471}
\acmYear{2026}
\acmJournal{PACMPL}
\acmVolume{10}
\acmNumber{OOPSLA2}
\acmArticle{339}
\acmMonth{10}
\acmSubmissionID{oopslab26main-p563-p}
\received{2026-03-17}
\received[accepted]{2026-08-06}

\begin{document}

\title{Semantics for 2D Rasterization}

\author{Bhargav Kulkarni}
\orcid{0009-0008-2320-6112}
\affiliation{%
  \institution{University of Utah}
  \city{Salt Lake City}
  \country{USA}
}
\email{bhargavk@cs.utah.edu}

\author{Henry Whiting}
\orcid{0009-0005-9559-8395}
\affiliation{%
  \institution{University of Utah}
  \city{Salt Lake City}
  \country{USA}
}
\email{u1598085@utah.edu}

\author{Pavel Panchekha}
\orcid{0000-0003-2621-3592}
\affiliation{%
  \institution{University of Utah}
  \city{Salt Lake City}
  \country{USA}
}
\email{pavpan@cs.utah.edu}

\begin{abstract}
Rasterization is the process of determining
  the color of every pixel drawn by an application.
Powerful rasterization libraries
  like Skia, CoreGraphics, and Direct2D
  put exceptional effort into drawing, blending, and rendering efficiently.
Yet applications are still hindered
  by the inefficient sequences of instructions
  that they ask these libraries to perform.
Even Google Chrome, a highly optimized web browser
  co-developed with the Skia rasterization library,
  still produces inefficient instruction sequences
  even on the top 100 most visited websites.
The underlying reason for this inefficiency
  is that rasterization libraries have complex semantics
  and opaque and non-obvious execution models.

To address this issue, we introduce \muskia,
  a formal semantics for the Skia 2D graphics library,
  and mechanize this semantics in Lean.
\muskia covers language and graphics features
  like canvas state, the layer stack, blending, and color filters,
  and the semantics itself is split into three strata
  to separate concerns and enable extensibility.
We then identify four patterns
  of sub-optimal Skia code produced by Google Chrome
  and write replacements for each pattern.
\muskia allows us to verify that the replacements are correct,
  including identifying numerous tricky side conditions.
We then develop a high-performance Skia optimizer
  that applies these patterns to speed up rasterization.
On \graphiteAppleNumBench Skia programs
  gathered from the top 100 websites,
  this optimizer yields a speedup of \graphiteAppleGeoSpeedup
  over Skia's most modern GPU backend,
  while taking just \graphiteAppleMaxOptTime for optimization.
The speedups persist across a variety of
  websites, Skia backends, and GPUs.
To provide true, end-to-end verification,
  optimization traces produced by the optimizer
  are loaded back into the \muskia semantics and translation-validated in Lean.
\end{abstract}




\maketitle

\section{Introduction}
\label{sec:intro}

Every pixel you see on a computer screen
  was produced by a process called \emph{rasterization}.
Programs like desktop, mobile, and web applications
  send a series of instructions to a \emph{rasterization library};
  these instructions draw shapes, render text,
  and blend overlapping visual elements.
Then, the rasterization library---%
  which might be Skia, CoreGraphics, Direct2D, Cairo, or something else---%
  executes the instructions,
  determining the color of every pixel on the screen.
Rasterization must be done at interactive rates,
  ideally 60 to 120 frames per second,
  to ensure a smooth experience for users.
Rasterization libraries are thus highly optimized.
Skia, for example, targets
  a wide variety of CPU- and GPU-based backends
  using both legacy and modern graphics APIs
  like OpenGL, Vulkan, and Metal.
In fact, between 2021 and 2025, the Skia team
  wrote an entirely new backend, Graphite,
  to improve rasterization times by about 15\%~\cite{graphite-blog-post}
  by leveraging new GPU APIs.
It's therefore no surprise
  that Skia is the rasterization library of choice for
  the Google Chrome, Firefox, and Ladybird browsers,
  the Android operating system, the Flutter mobile application framework,
  as well as for native applications like Sublime Text.

Despite this extreme performance focus,
  rasterization is \emph{not} a solved problem.
Web browsers, for example, struggle to render modern web pages,
  with trendy effects like partial transparency, animations, and blurs,
  at a stable 60 frames per second,
  especially on lower-end mobile devices.
The trend toward \SI{120}{Hz} displays only raises the bar.
The reason this is so hard is that
  while rasterization libraries implement individual instructions efficiently,
  \emph{clients ask them to execute inefficient instruction sequences}.
For example, the authors have manually examined
  the instructions that Chrome executes to raster the top 100 websites (by traffic)
  and identified numerous straightforward performance problems.
And Chrome is surely the most sophisticated Skia client:
  both are developed at the same company,
  with engineers working in such close coordination
  that Chrome and Skia make simultaneous releases.
Other Skia clients are even worse off.
Improving the quality of rasterization programs
  would dramatically reduce rasterization time,
  but doing that is unnecessarily difficult
  because rasterization libraries have strange, imperative semantics
  and an opaque, non-obvious execution model.

We address this problem with \muskia,
  a formal semantics for the Skia 2D rasterization library.
\muskia captures key features of the Skia API
  like canvas state, the layer stack, clipping, and blending.
More generally,
  since these features appear in all rasterization libraries,
  drawing from their common PostScript heritage,
  we see our formalization as providing a foundation
  for future language-driven rasterization research.
Our formalization is built in three strata---%
  a command language, a functional core, and an abstract model---%
  that separate concerns, simplify reasoning, and allow for extensibility.
The semantics is mechanized in Lean
  and enables automated reasoning and verification of \muskia programs.

To demonstrate the utility of \muskia,
  we identify four patterns of suboptimal Skia instructions
  generated by Google Chrome on the top 100 websites by traffic.
For each one, we provide a more efficient instruction sequence
  and prove the two sequences equivalent in Lean.
\muskia enables us to identify non-trivial side conditions
  and ensure that the optimization is correct.
We then build an optimizer that
  applies these rewrite rules before rasterization,
  enabling significant speed-ups to rasterization.
This simple optimizer results in an average speedup
  of \graphiteAppleGeoSpeedup over Skia's most modern Graphite backend
  on \graphiteAppleNumBench Chrome-generated Skia programs
  from the top 100 websites.
The optimizer is also highly efficient,
  with optimization taking at most \graphiteAppleMaxOptTime,
  and the optimizer improves rasterization time across GPUs and Skia back-ends.
Moreover, the optimizer is able to generate optimization traces
  that can be translation validated using \muskia,
  providing an end-to-end proof of correctness for those programs.

In short, the contributions in this paper are:
\begin{enumerate}
\item A new formalization of the Skia API and rasterization more generally (\Cref{sec:semantics})
\item A collection of focused optimizations proven valid using this formalization (\Cref{sec:rewrites})
\item A correct and efficient Skia optimizer that applies these optimizations (\Cref{sec:compiler}).
\end{enumerate}

\section{Skia}
\label{sec:skia}

\newcommand{\layerstack}[2]{%
  \begin{tikzpicture}[baseline=(base.center)]
    \node[anchor=south west] (base) at (0,0)
      {\frame{\includegraphics[scale=0.15]{#2}}};
    \path[use as bounding box] (base.south west) rectangle ([yshift=0.8em]base.north east);
    \node[anchor=south west] at (0.8em,0.8em)
      {\frame{\includegraphics[scale=0.15]{#1}}};
  \end{tikzpicture}%
}

Skia was originally an independent company
  founded by veterans of MacOS QuickDraw.
Google acquired Skia in 2005~\cite{skia-merger},
  open-sourced the library, and has maintained it since.
As of March 2026, its AUTHORS file lists 109 contributors,
  and it cuts stable milestone releases every four weeks.

A Skia client creates a ``surface''
  and then issues commands to that surface
  by calling methods of an associated \skcanvas object.
Then, when a \texttt{flush} method is called,
  the commands are executed to raster an image to the surface.%
\footnote{The CPU backend does execute commands eagerly,
  but most clients use the GPU backend.}
Commands can also be stored in an intermediate command buffer
  for serialization or replay.
We take this command language to be
  a stand-alone programming language
  and introduce its basic operations.

\subsection{Drawing and Painting}
\label{subsec:drawing-painting}

The \skdrawrect command draws a rectangle:%

\begin{figure}[H]
  \centering
  \begin{tabular}{c}
    \frame{\includegraphics[scale=0.15]{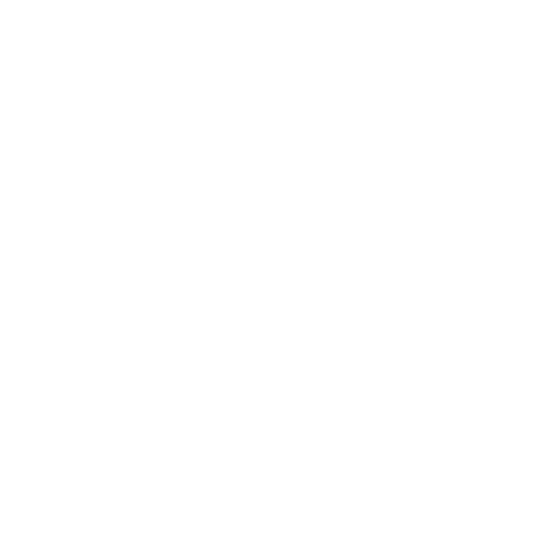}}
  \end{tabular}
  \quad
  \begin{tabular}{>{\smaller\ttfamily\raggedright\arraybackslash}m{0.42\linewidth}}
    SkPaint p;\\
    canvas->drawRect(rect1, p);
  \end{tabular}
  \quad
  \begin{tabular}{c}
    \frame{\includegraphics[scale=0.15]{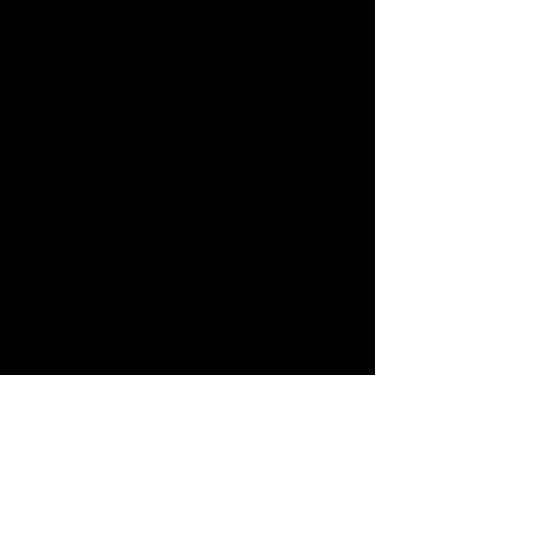}}
  \end{tabular}
  \Description{Empty canvas to the right, followed by a draw rect command, followed by a canvas with a black square}
\end{figure}


\noindent
Here, the layout is meant to evoke
  a Hoare triple:
  the left square shows the (empty) state of the canvas
  before the \skdrawrect command,
  while the right square shows the state of the canvas afterwards.
The first argument to \skdrawrect is the rectangle to be drawn,
  but the second argument, an \skpaint object, is more interesting.
An \skpaint describes \emph{how} a shape is drawn,
  for example how to \emph{fill} the shape.
The default fill for a \skpaint is the color black,
  but we can also change it to another color.

\begin{figure}[H]
  \centering
  \begin{tabular}{c}
    \frame{\includegraphics[scale=0.15]{drawing0.pdf}}
  \end{tabular}
  \quad
  \begin{tabular}{>{\smaller\ttfamily\raggedright\arraybackslash}m{0.42\linewidth}}
    SkPaint p;\\
    p.setColor(SK\_ColorRED);\\
    canvas->drawRect(rect1, p);
  \end{tabular}
  \quad
  \begin{tabular}{c}
    \frame{\includegraphics[scale=0.15]{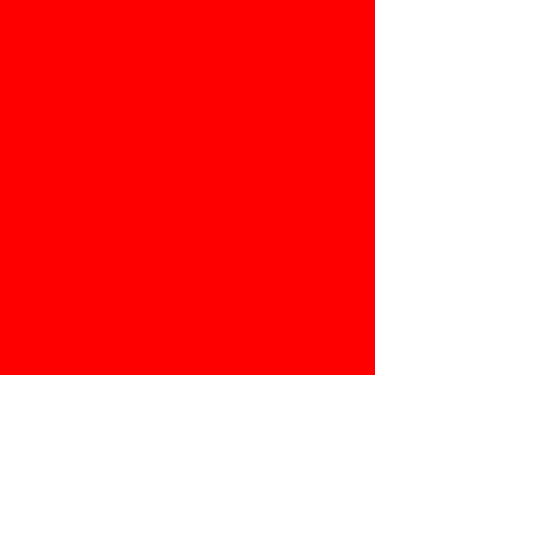}}
  \end{tabular}
  \Description{Empty canvas to the right, followed by a draw rect command, followed by a canvas with a red square}
\end{figure}

\noindent
It's also possible to specify a non-constant fill,
  like a gradient:

\begin{figure}[H]
  \centering
  \begin{tabular}{c}
    \frame{\includegraphics[scale=0.15]{drawing0.pdf}}
  \end{tabular}
  \quad
  \begin{tabular}{>{\smaller\ttfamily\raggedright\arraybackslash}m{0.42\linewidth}}
    SkPaint p;\\
    p.setShader(\\
    \quad SkShaders::LinearGradient(...));\\
    canvas->drawRect(rect1, p);
  \end{tabular}
  \quad
  \begin{tabular}{c}
    \frame{\includegraphics[scale=0.15]{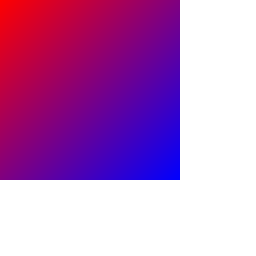}}
  \end{tabular}
  \Description{Empty canvas to the right, followed by a draw rect command, followed by a canvas with a gradient fill square}
\end{figure}

\noindent
Formally, filling is a stage of Skia's drawing pipeline,
  which assigns a color to every pixel in the shape.
More complex fills include images and shaders.

\subsection{Blending}
\label{subsec:blending}

Things get a little more interesting
  once there is more than one shape being drawn.
Suppose we draw a second, overlapping, blue square:









\begin{figure}[H]
  \centering
  \begin{tabular}{c}
    \frame{\includegraphics[scale=0.15]{drawing2.pdf}}
  \end{tabular}
  \quad
  \begin{tabular}{>{\smaller\ttfamily\raggedright\arraybackslash}m{0.42\linewidth}}
    p.setColor(SK\_ColorBLUE);\\
    canvas->drawRect(rect2, p);
  \end{tabular}
  \quad
  \begin{tabular}{c}
    \frame{\includegraphics[scale=0.15]{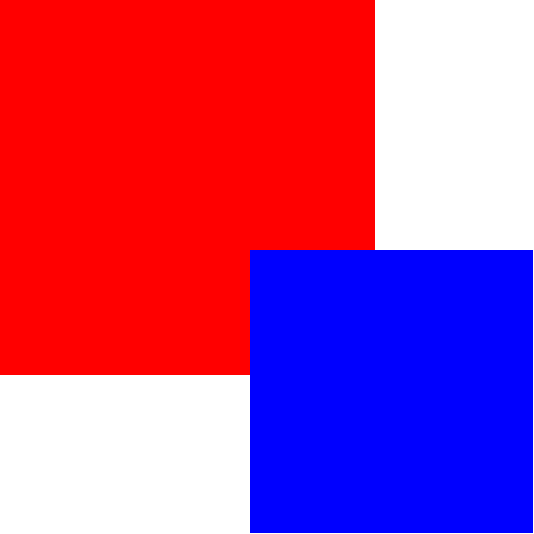}}
  \end{tabular}
  \Description{Canvas with a red square to the right, followed by a draw rect command, followed by a canvas with a red and blue square}
\end{figure}


Conceptually, drawing this second shape
  involves two stages.
First, as before, every pixel in \texttt{rect2}
  is assigned a fill color, here a solid blue.
But then, this assigned fill color is \emph{blended}
  with the existing color for that pixel.
That second stage is a bit difficult to see here,
  because the default \emph{blend mode},
  called \sksrcover, just chooses the new color.%
\footnote{In reality, \sksrcover blending is more complicated
  and depends on the opacity or \emph{alpha}
  of the new color,
  which can matter even for opaque fills due to anti-aliasing,
  but in our examples all the colors are opaque
  and we ignore anti-aliasing as a simplification.}
A program can, however, select a different blend mode instead.
For example, the \skmultiply blend mode %
  combines both colors being blended:

\begin{figure}[H]
  \centering
  \begin{tabular}{c}
    \frame{\includegraphics[scale=0.15]{drawing2.pdf}}
  \end{tabular}
  \quad
  \begin{tabular}{>{\smaller\ttfamily\raggedright\arraybackslash}m{0.42\linewidth}}
    p.setColor(SK\_ColorBLUE);\\
    p.setBlendMode(\\
    \quad SkBlendMode::kMultiply);\\
    canvas->drawRect(rect2, p);
  \end{tabular}
  \quad
  \begin{tabular}{c}
    \frame{\includegraphics[scale=0.15]{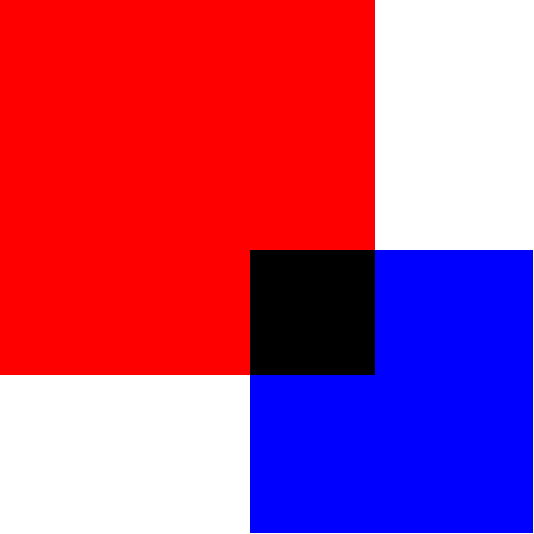}}
  \end{tabular}
  \Description{Canvas with a red square to the right, followed by a draw rect command, followed by a canvas with a red and blue square}
\end{figure}

In the intersection of the two rectangles,
  the initial red color is blended with the newly-drawn blue fill,
  which in the \skmultiply blend mode results in black.
Formally, blending is a stage in the Skia drawing pipeline
  that comes after filling and combines
  the fill color assigned to each pixel
  with the pre-existing color of that pixel.

\subsection{Clipping}
\label{subsec:clipping}

So far, \skdrawrect has been applying the fill
  to the entirety of the shape being drawn.
But Skia also provides a \skcliprect command
  to restrict or \emph{clip} the drawing bounds:%
\footnote{
  Skia allows clipping with arbitrary shapes,
  but rectangle and rounded-rectangle clips are especially fast,
  using scissor tests.}

\begin{figure}[H]
  \centering
  \begin{tabular}{c}
    \frame{\includegraphics[scale=0.15]{drawing0.pdf}}
  \end{tabular}
  \quad
  \begin{tabular}{>{\smaller\ttfamily\raggedright\arraybackslash}m{0.42\linewidth}}
    canvas->clipRect(clip\_rect);\\
    canvas->drawRect(rect1, redPaint);
  \end{tabular}
  \quad
  \begin{tabular}{c}
    \frame{\includegraphics[scale=0.15]{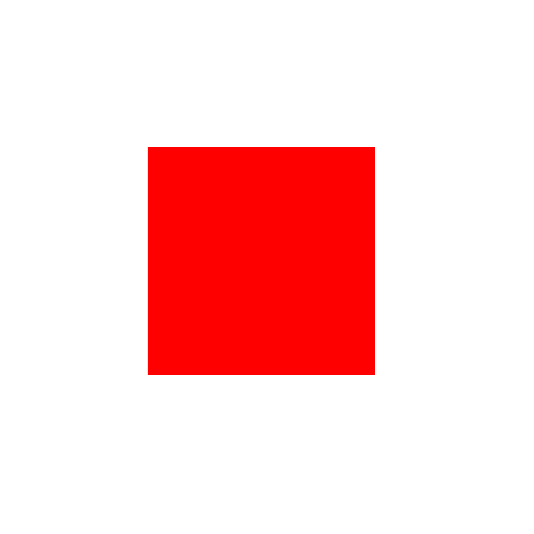}}
  \end{tabular}
  \Description{Empty canvas to the right, followed by a clip and draw rect command, followed by a canvas with a clipped red square}
\end{figure}

\noindent
Note that \skcliprect command is run first,
  before the \skdrawrect command.
In fact, \skcliprect does not modify the image at all;
  instead, it modifies \emph{canvas state}
  that influences later draw operations.
Formally, that influence happens in
  yet another stage of the Skia drawing pipeline
  that comes before filling and modifies the shape being drawn.

The canvas state is stored in the canvas object
  and can affect multiple drawing commands:
\begin{figure}[H]
  \centering
  \begin{tabular}{c}
    \frame{\includegraphics[scale=0.15]{drawing0.pdf}}
  \end{tabular}
  \quad
  \begin{tabular}{>{\smaller\ttfamily\raggedright\arraybackslash}m{0.42\linewidth}}
    canvas->clipRect(clip\_rect);\\
    canvas->drawRect(rect1, redPaint);\\
    canvas->drawRect(rect2, bluePaint);
  \end{tabular}
  \quad
  \begin{tabular}{c}
    \frame{\includegraphics[scale=0.15]{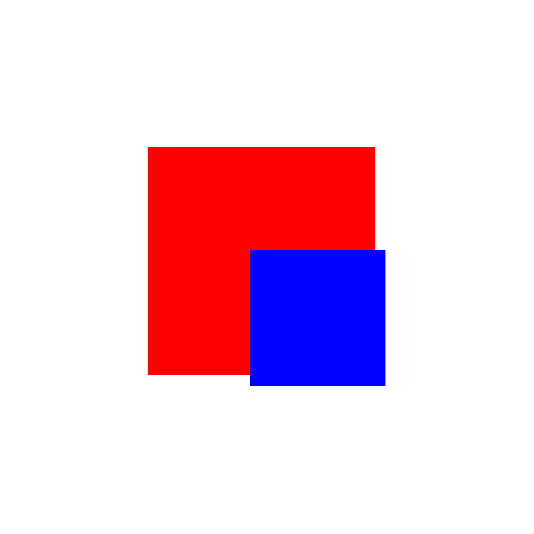}}
  \end{tabular}
  \Description{Empty canvas to the right, followed by a clip and two draw rect commands, followed by a canvas with a clipped red and blue square}
\end{figure}

Naturally, one might want to limit
  which drawing commands a clip command applies to.
Luckily, Skia maintains a \emph{stack} of canvas states;
  \skcliprect affects the top of the stack,
  which is used by drawing commands,
  while \sksave and \skrestore commands
  push and pop from the stack:

\begin{figure}[H]
  \centering
  \begin{tabular}{c}
    \frame{\includegraphics[scale=0.15]{drawing0.pdf}}
  \end{tabular}
  \quad
  \begin{tabular}{>{\smaller\ttfamily\raggedright\arraybackslash}m{0.42\linewidth}}
    canvas->save();\\
    canvas->clipRect(clip\_rect);\\
    canvas->drawRect(rect1, redPaint);\\
    canvas->restore();\\
    canvas->drawRect(rect2, bluePaint);
  \end{tabular}
  \quad
  \begin{tabular}{c}
    \frame{\includegraphics[scale=0.15]{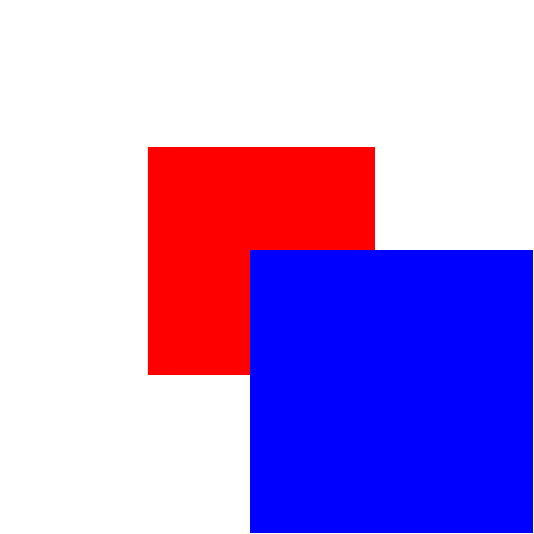}}
  \end{tabular}
  \Description{Empty canvas to the right, followed by a clip and and draw rect command out inside a save command and one draw rect command outside, followed by a canvas with a clipped red and unclipped blue square}
\end{figure}

\noindent
In other words, \sksave and \skrestore act like brackets,
  restricting the effect of \skcliprect to the enclosed commands.

\subsection{Layers}
\label{subsec:layers}

Besides the stack of canvas states modified by \sksave and \skrestore,
  there is a stack of drawing surfaces
  and \sksavelayer and \skrestore commands that push and pop from it.
Conceptually, \sksavelayer allocates
  a new, empty drawing surface:

\begin{figure}[H]
  \centering
  \begin{tabular}{c}
    \frame{\includegraphics[scale=0.15]{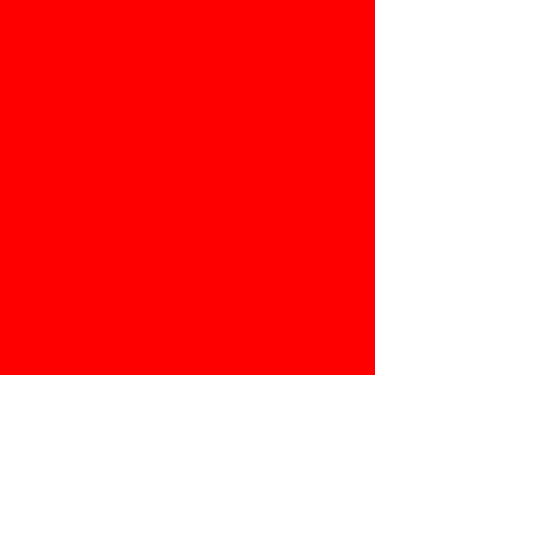}}
  \end{tabular}
  \quad
  \begin{tabular}{>{\smaller\ttfamily\raggedright\arraybackslash}m{0.42\linewidth}}
    canvas->saveLayer();
  \end{tabular}
  \quad
  \layerstack{drawing0.pdf}{drawing2.pdf}
  \Description{Canvas with red square, followed by a save layer command, followed by a stack of a red square canvas and empty square}
\end{figure}

All drawing commands until the corresponding \skrestore
  draw to this new surface.

\begin{figure}[H]
  \centering
  \layerstack{drawing0.pdf}{drawing2.pdf}
  \quad
  \begin{tabular}{>{\smaller\ttfamily\raggedright\arraybackslash}m{0.42\linewidth}}
    canvas->drawRect(rect2, bluePaint);
  \end{tabular}
  \quad
  \layerstack{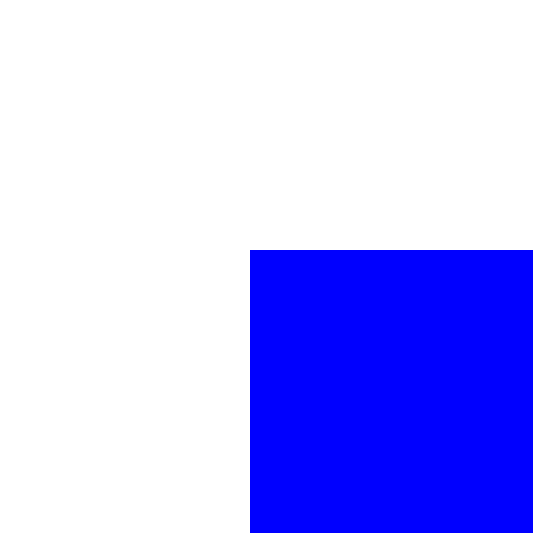}{drawing2.pdf}
  \Description{Stack of a red square canvas and empty canvas, followed by a draw rect command, followed by a stack of a red square canvas and blue square canvas}
\end{figure}

Then, when the \skrestore command is reached,
  the whole resulting drawing surface is blended back into
  the original drawing surface.

\begin{figure}[H]
  \centering
  \layerstack{drawing9.pdf}{drawing2.pdf}
  \quad
  \begin{tabular}{>{\smaller\ttfamily\raggedright\arraybackslash}m{0.42\linewidth}}
    restore();
  \end{tabular}
  \quad
  \begin{tabular}{c}
    \frame{\includegraphics[scale=0.15]{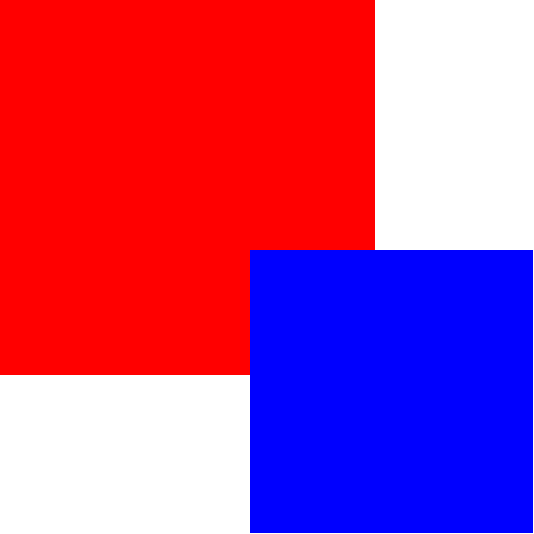}}
  \end{tabular}
  \Description{Stack of a red square canvas and blue square canvas, followed by a restore command, followed by a canvas with a red and blue square}
\end{figure}

The \sksavelayer operation accepts an \skpaint argument
  that can change the blend used during \skrestore.
For example, with \skmultiply blending,
  the previous example would look like so:

\begin{figure}[H]
  \centering
  \layerstack{drawing9.pdf}{drawing2.pdf}
  \quad
  \begin{tabular}{>{\smaller\ttfamily\raggedright\arraybackslash}m{0.42\linewidth}}
    restore(); // multiply blending
  \end{tabular}
  \quad
  \begin{tabular}{c}
    \frame{\includegraphics[scale=0.15]{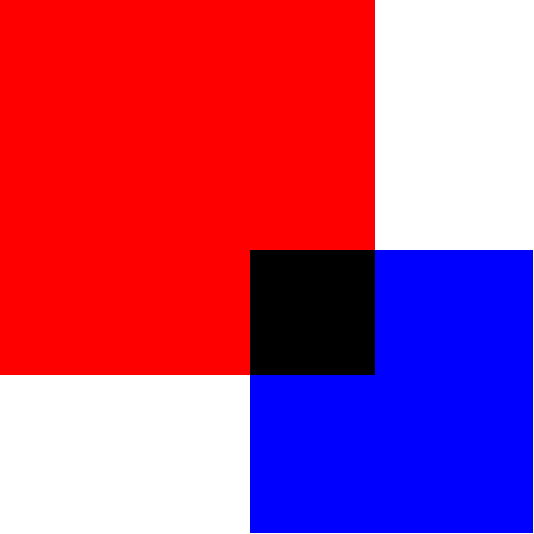}}
  \end{tabular}
  \Description{Stack of a red square canvas and blue square canvas, followed by a restore command, followed by a canvas with a red and blue square}
\end{figure}

It is important to note that \sksavelayer, besides allocating a new drawing surface, also pushes
  onto the canvas stack like \sksave does.

\subsection{Masking}
\label{subsubsec:masking}

One prominent use for \sksavelayer is masking.
This refers to clipping one image
  based on the contents of another image,
  and it can be achieved using \sksavelayer
  with \skdstin blending:

\begin{figure}[H]
  \centering
  \begin{tabular}{c}
    \frame{\includegraphics[scale=0.15]{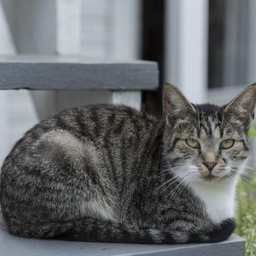}}
  \end{tabular}
  \quad
  \begin{tabular}{>{\smaller\ttfamily\raggedright\arraybackslash}m{0.42\linewidth}}
    SkPaint blend\_p;\\
    blend\_p.setBlendMode(\\
    \quad SkBlendMode::kDstIn);\\
    canvas->saveLayer(blendPaint);
  \end{tabular}
  \quad
  \layerstack{drawing0.pdf}{cat.png}
  \Description{Canvas with a cat, followed by a save layer command, followed by a stack of a cat canvas and a empty canvas}
\end{figure}

Here the initial state is the image%
\footnote{%
Photograph by Carol M.
Highsmith, ``A contented cat at Yoder's Amish Home, an authentic Amish farm near Walnut Creek,
Ohio.''
Public domain, Library of Congress.~\cite{cat}} to be masked,
  while the new layer represents the mask.
We now draw the mask shape on the new layer.

\begin{figure}[H]
  \centering
  \layerstack{drawing0.pdf}{cat.png}
  \quad
  \begin{tabular}{>{\smaller\ttfamily\raggedright\arraybackslash}m{0.42\linewidth}}
    SkPaint p;\\
    canvas->drawRect(rect1, p);\\
    canvas->drawRect(rect2, p);
  \end{tabular}
  \quad
  \layerstack{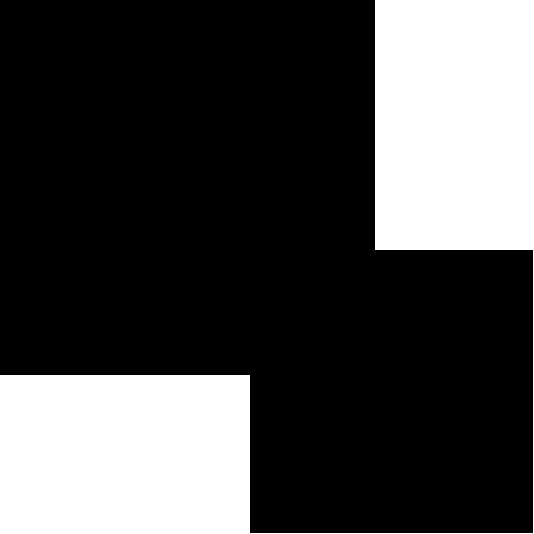}{cat.png}
  \Description{Stack of a cat canvas and an empty canvas, followed by two draw rect commands, followed by a stack of a cat canvas and a canvas with 2 rectangles}
\end{figure}

Finally, on \skrestore,
  the initial image and the mask are combined
  using \skdstin blending.
\skdstin is very simple:
  it simply keeps all pixels from the bottom layer
  that are colored in in the top layer.%
\footnote{
  The behavior with opacity is slightly more complex,
  but the same basic idea remains.
}

\begin{figure}[H]
  \centering
  \layerstack{mask.pdf}{cat.png}
  \quad
  \begin{tabular}{>{\smaller\ttfamily\raggedright\arraybackslash}m{0.42\linewidth}}
    restore();
  \end{tabular}
  \quad
  \begin{tabular}{c}
    \frame{\includegraphics[scale=0.15]{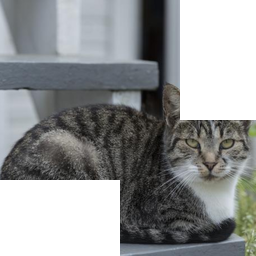}}
  \end{tabular}
  \Description{Stack of a cat canvas and a canvas with 2 rectangles, followed by a restore, followed by a clipped cat canvas}
\end{figure}

\skdstin blending is far more powerful than \skcliprect
  because it also allows masking with
  arbitrary Skia-drawable images.
For example,
  here is the previous image,
  masked with a radial gradient.

\begin{figure}[H]
  \centering
  \layerstack{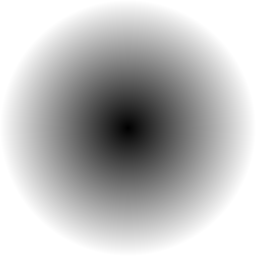}{cat.png}
  \quad
  \begin{tabular}{>{\smaller\ttfamily\raggedright\arraybackslash}m{0.42\linewidth}}
    restore();
  \end{tabular}
  \quad
  \begin{tabular}{c}
    \frame{\includegraphics[scale=0.15]{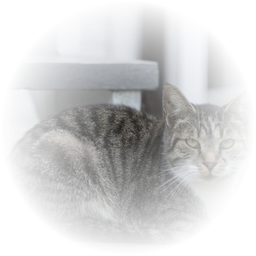}}
  \end{tabular}
  \Description{Stack of a cat canvas and a canvas with a gradient, followed by a restore, followed by a clipped cat canvas}
\end{figure}

\subsection{Equivalences}
\label{subsec:equivalences}

Nonetheless, simple rectangular clips
  can be achieved using either \skcliprect
  or with a \sksavelayer with \skdstin blending:

\begin{figure}[H]
  \centering
  \begin{tabular}{c}
    \frame{\includegraphics[scale=0.15]{drawing0.pdf}}
  \end{tabular}
  \quad
  \begin{tabular}{l}
    \begin{tabular}{>{\smaller\ttfamily\raggedright\arraybackslash}m{0.42\linewidth}}
      canvas->clipRect(rect1);\\
      canvas->drawImage(...);
    \end{tabular}\\
    \noalign{\vskip 1.4em}
    \cline{1-1}
    \noalign{\vskip 1.4em}
    \begin{tabular}{>{\smaller\ttfamily\raggedright\arraybackslash}m{0.42\linewidth}}
      canvas->drawImage(...);\\
      SkPaint b;\\
      b.setBlendMode(SkBlendMode::kDstIn);\\
      canvas->saveLayer(b);\\
      SkPaint p;\\
      canvas->drawRect(rect1, p);\\
      canvas->restore();
    \end{tabular}
  \end{tabular}
  \quad
  \begin{tabular}{c}
    \frame{\includegraphics[scale=0.15]{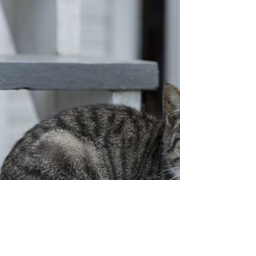}}
  \end{tabular}
  \Description{Empty canvas, followed by a 2 equivalent Skia programs, followed by a clipped cat canvas}
\end{figure}

However, the \sksavelayer-based program
  is far less efficient than the \skcliprect-based one.
It has to allocate a whole new layer,
  draw additional pixels to it,
  and blend the results together,
  while the \skcliprect-based one uses a fast path.
In fact, it is common for the same image
  to be achievable with many different Skia programs,
  some of which are much more efficient than others.
This motivates the rest of the paper:
\Cref{sec:semantics} develops a semantics,
  which allows proving that two Skia programs have identical results.
\Cref{sec:rewrites} uses that semantics
  to verify several rewrite rules over Skia programs.
And \Cref{sec:compiler} then uses those rewrites
  to build a general-purpose Skia optimizer.
\Cref{sec:eval} demonstrates that this optimizer results
  in dramatically faster rasterization
  of a wide range of real-world Skia programs.

\section{Semantics}
\label{sec:semantics}

Our Skia semantics consists of three strata.
The bottom-most stratum is an abstract model of
  images, shapes, blends, and filters defined in terms
  of abstract types \Point and \Color,
  which can be instantiated to model
  as much of the actual Skia API as necessary.
The middle stratum is functional operations
  \DrawShape and \BlendLayer
  parameterized by a \Paint object,
  all of which denote to our abstract model.
The top stratum is an imperative language, \muskia,
  whose operational semantics provides a model of Skia.

\subsection{Abstract Model}
\label{subsec:abstractmodel}

\begin{figure}
  \centering
  \begin{tabular}{L L L}
    Image  & : \; Point $\to$ Color & \textrm{\textit{e.g.}}\; {\sffamily
                                      Solid($c : \text{Color}$), LinearGradient($\ldots$),
                                      $\ldots$} \\[4pt]
    Shape  & : \; Point $\to$ Bool & \textrm{\textit{e.g.}}\; {\sffamily
                                     Rect($\ldots$), RRect($\ldots$),
                                     Path($\ldots$), Text($\ldots$)} \\[4pt]
    Blend  & : \; Color $\times$ Color $\to$ Color & \textrm{\textit{e.g.}}\;
                                                     {\sffamily SrcOut, DstIn,
                                                     $\ldots$} \\[4pt]
    Filter & : \; Color $\to$ Color & \textrm{\textit{e.g.}}\; {\sffamily Id,
                                      Luma, $\ldots$}
  \end{tabular}
  \caption{%
    Our abstract model of \Image{}s
      and \Shape{}s is built from
      abstract types \Point{}s and \Color{}s.
    We also include the various instantiations of the types of the abstract model we use in
      our formalization.
  }
  \Description{Abstract model of Images and Shapes, along with instantiations of the same}
\label{fig:model}
\end{figure}

In traditional programming language semantics,
  the abstract model---%
  pairs, functions, integers, and the like---%
  is implicit and well understood.
In the graphics domain, the core types
  are not so well established,
  so the abstract model must be explicitly defined.
The abstract model for \muskia
  is based around two core abstract types,
  \Point and \Color.
The \Point and \Color types
  are then used to define additional types,
  as shown in \Cref{fig:model}.
A special \Transparent color is also assumed.
An \Image is defined
  as a \Color for every \Point
  and \Shape{}s are defined
  as predicates on \Point{}s.
Equivalence of two \Image{}s is defined extensionally:
  the same \Color for each \Point.
Additionally, since a major focus of this semantics
  is reasoning about color blending and filtration,
  \Blend{}s and \Filter{}s are defined
  as binary and unary operators on colors.

An instantiation of the abstract model
  would then define specific \Image{}s
  (solid color fills, gradients, image files),
  \Shape{}s (rectangles, paths, text outlines),
  \Blend{}s (roughly two dozen),
  and \Filter{}s (a few built-in, but extensible with shaders),
  along with equivalences between them.
\Cref{fig:model} shows some of the objects of each type.
This provides extensibility for \muskia:
  an instantiation of the abstract model
  can define more or fewer of these objects,
  and provide more or fewer theorems about them,
  allowing a trade-off between the accuracy and complexity
  of our formalism.
Our Lean mechanization, for example,
  treats most shapes as entirely symbolic,
  with only minimal support for reasoning about
  subset relations between them,
  sufficient for the rewrites of \Cref{sec:rewrites}
  and the optimizer of \Cref{sec:compiler}.
Future work may expand the abstract model
  or more carefully model its objects
  so as to provide more theorems about them.

The abstract model also provides some flexibility
  for choosing either simpler or more accurate
  models of \Point and \Color.
The actual Skia implementation
  defines \Point to be a pair of 32-bit floats
  and \Color to be a 32-bit unsigned integer
  interpreted as ARGB pre-multiplied byte-sized color,%
\footnote{
``ARGB'' gives the in-memory byte order
  of the alpha, red, green, and blue components of a color.
The interpretation of these bytes as an unsigned 32-bit integer
  will depend on the endianness of the platform itself.
The colors are premultiplied, meaning
  that a color with an alpha value of $\alpha$
  will have its red, green, and blue components
  range from $0$ to $\alpha$;
  this form makes blending computations somewhat more efficient.
}
  and operations on points and colors
  thus reflect effects from anti-aliasing and rounding.
The exact anti-aliasing and rounding behavior
  is not guaranteed by Skia
  and is also exceptionally complex,
  including hardware dependence in some back-ends.
Our Lean formalization instead uses an abstract model
  where points and colors are represented by real numbers,
  which in practice means it is accurate
  \emph{up to} anti-aliasing and rounding.

\subsection{Skia as Layers}
\label{subsec:sem-layers}

\begin{figure}
  \begin{minipage}{0.51\linewidth}
    \centering
    \begin{tabular}{L}
      Layer ::= \\
      | Empty() \\
      | DrawShape($\ell: \Layer$, $s: \Shape$, $p: \Paint$) \\
      | BlendLayer($\ell_{\textsf{bot}}: \Layer$, $\ell_{\textsf{top}}: \Layer$, $p: \Paint$) \\[4pt]

      Paint ::= \\
      | Paint(\textit{fill} : Image, \textit{filter} : Filter, \textit{blend} : Blend)
    \end{tabular}
    \caption{
      \Layer objects store images and are modified
        by drawing and blending operations.
      Each drawing or blending operation
        is parameterized by a \Paint object.
      \Layer terms denote to the abstract model
        of \Cref{subsec:abstractmodel}.
      }
    \label{fig:layerlang}
  \end{minipage}%
  \hfill%
  \begin{minipage}[T]{0.45\linewidth}
    \centering
    \begin{tabular}{L}
      Command ::= \\
      | Draw($g: \Shape$, $p : \Paint$) \\
      | Clip($m : \Shape$) \\
      | Save(); $c$; Restore() \\
      | SaveLayer($p: \Paint$); $c$; Restore() \\
      | $c$; $c$
    \end{tabular}
    \caption{The \muskia command language.
      Note that \Save/\Restore and \SaveLayer/\Restore pairs
        are defined as structured terms
        wrapping a subsequence of commands.
      }
    \label{fig:muskia}
  \end{minipage}
  \Description{Definitions of the Layer and muSkia language}
\end{figure}

The execution model of Skia centers around layers,
  which store images
  and are modified by drawing and blending.
\Cref{fig:layerlang} thus introduces \Layer objects
  with \DrawShape and \BlendLayer operations.
\DrawShape draws a single shape $s$ on a \Layer $\ell$
  yielding the new state of $\ell$,
  while \BlendLayer blends the layer $\ell_\textsf{top}$
  into $\ell_\textsf{bot}$, yielding the new state of $\ell_\textsf{bot}$.
Each operation is parameterized by a \Paint object,
  which defines a blend mode, fill type, and color filter.
The \Paint object configures the various phases
  of the Skia drawing pipeline.
Because \Layer has only these two constructors
  (plus the special \Empty layer),
  it is particularly easy to induct on in proofs.

\begin{figure}
  \begin{align*}
    \denote{\Empty()}(\textsf{pt})
    &= \Transparent\\[2pt]
    \denote{\BlendLayer(\ell_\textsf{bot}, \ell_\textsf{top}, p)}(\mathsf{pt})
    &= \mathsf{blend}_p\left(\denote{\mathsf{\ell_{bot}}}(\mathsf{pt}),\;
        \textsf{filter}_p\left(\denote{\mathsf{\ell_{top}}}(\mathsf{pt})\right)
      \right) \\[2pt]
    \denote{\DrawShape(\ell, s, p)}(\mathsf{pt})
    &= \mathsf{blend}_p\left(\denote{\ell}(\mathsf{pt}),\;
      \textsf{filter}_p\left(
       \left\{
       \begin{array}{ll}
         \mathsf{fill}_p(\mathsf{pt}) & s(\mathsf{pt}) \\
         \Transparent & \textrm{else}
       \end{array}
       \right\}
       \right)\right)\\
  \end{align*}
  \caption{Denotational semantics for \Layer terms.}
  \label{fig:denote}
  \Description{Denotational semantics for Layer terms}
\end{figure}

The \layer constructors denote to the abstract model,
  as shown in \Cref{fig:denote}.
\Empty has a straightforward denotation
  to an all-\Transparent image.
\BlendLayer is likewise straightforward,
  first applying an image filter to the top layer
  and then blending it with the bottom layer,
  both pointwise and with
  the filter and blend mode determined by the \Paint object.
\DrawShape is the most complex.
Conceptually, it first computes which points
  are within the shape $s$;
  then computes the fill color for each one;
  and finally blends the layer's current color
  and computed color for each point.
The blend is performed even for points outside $s$;
  this is critical for, for example, \textsf{DstIn} blending.

\Layer{}s are equivalent up to denotation,
  and permit substitution.
That is, if $\denote{\ell} = \denote{\ell'}$,
  then $\denote{m} = \denote{m[\ell/\ell']}$.
This fact means that equivalences over \Layer terms
  can be used as rewrite rules,
  which we leverage in \Cref{sec:rewrites}
  to prove Skia optimizations correct.

\subsection{\texorpdfstring{\muskia}{muSkia}}
\label{subsec:muskia}

\muskia is a simplified version of the Skia API itself.
A \muskia program consists
  of a sequence of commands
  (see \Cref{fig:muskia}).
These commands include
  all the features described in \Cref{sec:skia},
  including the \Draw command for drawing,
  the \Clip command for manipulating the clip state,
  and the \Save, \SaveLayer, and \Restore
  stack manipulation commands.
\muskia diverges from the Skia API
  in two main ways.
First, the syntax definition is more regular
  to eliminate unimportant special cases:
  \Save, \SaveLayer, and \Restore
  must be properly bracketed,
  while the \Draw and \Clip commands
  take arbitrary \Shape{}s,
Second, only a small fraction of the API is modeled.
Our Lean formalization extends the grammar in \Cref{fig:muskia}
  with several additional features,
  including opacity, styles, and transforms,
  and we expect adding even more features to be straightforward.

\muskia programs manipulate a state $\Sigma$:
\[
\Sigma:\List[\Layer] \times \List[\Shape]
\]
The first list in $\Sigma$ is the stack of layers
  manipulated by \SaveLayer,
  while the second list is the stack of canvas states
  manipulated by \Save.
In the minimal \muskia language of \Cref{fig:muskia},
  the canvas state is a single \Shape
  because the only canvas state operation is \Clip;
  our Lean formalization adds additional fields.
The initial state is
  $\langle [ \texttt{Empty()} ], [ \mathcal{U} ] \rangle$,
  an \Empty layer and the full clip state,
  each as single-element stacks;
  we treat the head of the list
  as the top of the stack.

A \muskia program then performs
  a sequence of steps  $c, \Sigma \longmapsto \Sigma'$.
The simplest is the \textsf{Draw} command,
  which modifies the top \textsf{Layer}:
\begin{mathpar}
  \inferrule
  { }
  {
    \Draw(g, p),
    \langle \ell :: L, m :: C \rangle
    \;\longmapsto\;
    \langle \DrawShape(\ell, g \cap m, p) :: L, m :: C \rangle
  }
\end{mathpar}
\noindent
Note that the \DrawShape operation
  is passed the intersection of the shape $g$ being drawn
  and the current clip mask $m$.

The \textsf{Clip} command modifies the canvas state,
  intersecting it with the new clip geometry:
\begin{mathpar}
  \inferrule
  { }
  {
    \Clip(m'),
    \langle L, m :: C\rangle
    \;\longmapsto\;
    \langle L, (m \cap m') :: C \rangle
  }
\end{mathpar}
\noindent
Note that the actual Skia \texttt{clipRect} command
  has an optional \texttt{SkClipOp} parameter
  that allows taking the difference instead of intersecting
  the new mask with the clip state;
  this paper does not show that (straightforward) extension
  but our Lean formalization does include it.

The \textsf{Save} command
  duplicates the top canvas state,
  runs the bracketed commands,
  and then discards the new canvas state
  when its matching \textsf{Restore} is reached:
\begin{mathpar}
  \inferrule
  {
    c,
    \langle L, m :: m :: C \rangle
    \;\longmapsto\;
    \langle L', m' :: m :: C \rangle
  }
  {
    (\Save(); c; \Restore()),
    \langle L, m :: C\rangle
    \;\longmapsto\;
    \langle L', m :: C \rangle
  }
\end{mathpar}
\noindent
Note that \Restore does not make
  any changes to the layer state.

The \textsf{SaveLayer} command, meanwhile,
  allocates a new, \Empty layer
  to push onto the layer stack and,
  when the \textsf{Restore} command is reached,
  blends that new layer with the one below:
\begin{mathpar}
  \inferrule
  {
    c,
    \langle \Empty() :: \ell :: L, m :: m :: C \rangle
    \;\longmapsto\;
    \langle \ell' :: \ell :: L, m' :: m :: C \rangle
  }
  {
    (\SaveLayer(p); c; \Restore()),
    \langle \ell :: L, m :: C\rangle
    \;\longmapsto\;
    \langle \BlendLayer(\ell, \ell', p) :: L, m :: C \rangle
  }
\end{mathpar}
\noindent
The inference rule for sequential composition is straightforward
  and, along with various extensions to \muskia,
  is given in our Lean formalization.

The \muskia semantics is deterministic,
  and programs never get stuck.%
\footnote{
Proving this requires proving that Skia programs only ever
  modify the top of the layer and canvas stack,
  and do not change the stack size.
This is proven by induction.
}
The final state consists of a single layer,
  whose denotation is the program's output.

\subsection{Lean Formalization}
\label{subsec:lean}

We have mechanized all three strata of this formalization---%
  the abstract model, \Layer terms, and \muskia commands,
  and their respective denotational / operational semantics,
  in the Lean~4 interactive theorem prover~\cite{lean4}.
Our mechanization, which we submitted as an artifact,
  is under a thousand lines of code.
We also provide an instantiation of the \Color abstract type
  as a 4-tuple of real numbers
  representing premultiplied ARGB colors.
This means that the alpha channel $\alpha$ is bounded between $0$ and $1$,
  while the three color channels are bounded between $0$ and $\alpha$.
By formalizing \Color with reals,
  equivalences in this abstract model are only up to rounding.
We have discussed these choices with the Skia engineering team,
  who confirmed that rounding behavior,
  while important, is acceptable to change during optimization.

Our formalization also includes a number of extensions
  not discussed in this section.
We formalize ``style'', like fill versus stroke,
  as a unary function on \Shape{}s,
  and handle opacity accurately during blending,
  which requires care with certain blend modes.
We also extend the canvas state with an affine transform,
  manipulated by translate, rotate, scale, and perspective operations,
  and add \texttt{SkClipOp}-style clip operations.
All these extensions involve new \muskia commands,
  or new parameters to existing \muskia commands,
  but do not affect the \Layer terms.
This makes it easy to write proofs that remain valid
  as the semantics are extended.
More speculatively, we also expect that
  \Layer terms will be useful for formalizing
  rasterization libraries besides Skia,
  since most existing rasterization libraries
  share a similar drawing pipeline
  due to a shared heritage from PostScript.

There are still some parts of Skia's API
  that our Lean mechanization does not cover,
  like mask filters, blurs, and image filters,
  as well as even more obscure portions of the API
  like the SkSL shading language or support for bitmap images.
These have little effect on the rewrite rules of \Cref{sec:rewrites},
  but we do expect that they could be added to \muskia if a need arose.

\section{\texorpdfstring{Rewrites in \muskia}{Rewrites in muSkia}}
\label{sec:rewrites}

\muskia enables rigorous, verified reasoning
  about the equivalence of Skia programs.
As a case study,
  we collected Chrome-generated Skia programs
  from the top 100 websites, including popular web pages like
  Amazon, Github, New York Times, and Tiktok.%
\footnote{
And many international pages like Bilibili, Mail.ru, and Yahoo.jp.
}
A manual examination showed that many included
  suboptimal patterns that used more \SaveLayer{}s than necessary,
  which means rasterization allocated unnecessary GPU memory
  and performed unnecessary blends.
We identified four common patterns of suboptimal Skia instructions;
  some patterns were common across many pages,
  while others were specific to particular web frameworks.
For each one, we wrote a more optimized pattern
  and verified each replacement using the \muskia semantics,
  including identifying the necessary side conditions
  and proving the necessary equivalences for our real abstract model.

Formally, two \muskia programs are equal
  when they evaluate to the same \Image in the \muskia semantics.
Because \muskia programs are imperative,
  equivalence does \emph{not} generally allow substitution.
However, because \muskia commands
  only modify the top of the state or layer stack,
  substitution directly inside \SaveLayer \emph{is} valid.%
\footnote{
We mechanize this theorem in Lean.
}
The common patterns we identify
  are therefore whole-program or whole-layer patterns.

\subsection{SrcOver \SaveLayer{}s}
\label{subsec:src-over-layer}

\SrcOver is by far the most common blend mode,
  corresponding to the normal intuitive meaning of
  ``drawing over'' some other image,
  and \SaveLayer{}s with \SrcOver blend mode
  are often unnecessary,
  suggesting the following idealized equivalence:
\begin{center}
  \begin{tabular}{L c L}
    \begin{tabular}{@{}l@{}}
      $\ell_1\ldots$ \\
      SaveLayer($p$ \{ \SrcOver, \Id \}); \\
      \quad $\ell_2\ldots$ \\
      Restore(); \\
      $\ell_3\ldots$
    \end{tabular}
    & = &
    \begin{tabular}{@{}l@{}}
      $\ell_1\ldots$ \\
      Save(); \\
      \quad $\ell_2\ldots$ \\
      Restore(); \\
      $\ell_3\ldots$
    \end{tabular}
  \end{tabular}
\end{center}
\noindent
Here, the syntax $\ell \ldots$ means
  any sequence of Skia commands,
  while the syntax $p \{ \SrcOver, \Id \}$ means
  any \Paint object $p$ with a \SrcOver blend mode
  and \Id color filter.

We prove that this equivalence holds under the condition
  that all drawing commands in $\ell_2$ use \SrcOver blend mode
  and further that $\ell_2$ contains no \SaveLayer commands.%
\footnote{
Our Lean formalization also includes opacity,
  so additionally restricts the paint $p$ to be opaque.
}
This restriction is necessary to ensure
  that \DstIn or other blends performed in $\ell_2$
  are properly isolated from $\ell_1$.
In the replacement pattern on the right hand side,
  the \SaveLayer commands are replaced with \Save.
This ensures that changes to canvas state in $\ell_2$
  are properly isolated from $\ell_3$,
  while still avoiding the extra layer allocation and blend
  performed by \SaveLayer.

This equivalence is proven by induction over $\ell_2$.
It is convenient to assume
  that sequential composition is left-associated;
  induction then proceeds by ``peeling off''
  the last drawing command in $\ell_2$
  and case-splitting on it.
After expanding out the \muskia semantics,
  the inductive step requires proving
  the following equivalence of \Layer terms:
  \[
      \BlendLayer(\ell_1, \DrawShape(\ell_2, g, p_d \{ \SrcOver \}), p)
      =
      \DrawShape(\BlendLayer(\ell_1, \ell_2, p), g, p_d)
  \]
Note that, per the restriction of $\ell_2$,
  the \Draw command uses a \SrcOver blend.
Applying this rewrite once ``peels'' a single \Draw command
  out of the inner \SaveLayer,
  and applying it repeatedly results, inductively,
  in the intended equivalence.

The \Layer equivalence, in turn,
  boils down to the following equivalence in our abstract model:
\[
  \SrcOver(a, \SrcOver(b, c)) = \SrcOver(\SrcOver(a, b), c)
\]
We prove this equivalence directly
  for our real-valued instantiation of the abstract model
  using the standard definition of \SrcOver blending.

\subsection{Dstin to Clip}
\label{subsec:dstin-clip}

In a number of cases, Chrome uses \dstin
  for general-purpose masking,
  even though using a \Clip command would be
  equivalent and more efficient.
This suggests the following idealized equivalence:

\begin{center}
  \begin{tabular}{L c L}
    \begin{tabular}{@{}l@{}}
      $\ell_1\ldots$ \\
      SaveLayer($p_1$ \{ \DstIn \}); \\
      \quad $\ell_2\ldots$ \\
      \quad Draw($g$, $p_2$ \{ \textsf{Solid}($c_2$), \SrcOver \}); \\
      Restore()
    \end{tabular}
    & = &
    \begin{tabular}{@{}l@{}}
      Clip($g$); \\
      $\ell_2\ldots$ \\
      $\ell_1\dots$ \\
    \end{tabular}
  \end{tabular}
\end{center}

We prove that this equivalence holds under the condition
  that $\ell_1$ does not contain any \SaveLayer operations,
  that $\ell_2$ contains \emph{only} \Clip operations,%
\footnote{
Our Lean formalization also includes \texttt{SkClipOp}-style
  clip operations so further restricts $\ell_2$ \Clip operations
  to only use the intersection operator.
}
  and that the color $c_2$ is opaque
  after applying $p_2$'s and then $p_1$'s color filters.

The challenge with verifying this rewrite
  is that \Clip operations in $\ell_2$
  do not apply to the operations in $\ell_1$ on the left hand side,
  but do apply to operations in $\ell_1$ on the right hand side.
The equivalence holds because
  the $\ell_2$ \Clip{}s intersect with $g$,
  which then acts as a mask for the entire image drawn by $\ell_1$.
The proof therefore proceeds by induction over $\ell_1$,
  and we define a helper function \textsf{clipAll} on \Layer terms to aid the proof:
\begin{align*}
  &\mathsf{clipAll} : \Layer \to \Shape \to \Layer \\
  &\mathsf{clipAll}(\Empty(), m) = \Empty() \\
  &\mathsf{clipAll}(\Draw(\ell, g, p), m)
                   = \Draw(\mathsf{clipAll}(\ell, m), g \cap m, p)
\end{align*}
Note that \textsf{clipAll} is not defined
  on \BlendLayer{}s,
  which is fine because we assume that $\ell_1$
  does not contain \SaveLayer{}s;
  in Lean, \texttt{clipAll} returns a dummy layer in that case.
We then prove, by induction,
  that commuting \Clip over $\ell$ results in $\mathsf{clipAll}(\ell)$.

With this induction proven,
  the core \Layer equivalence becomes:
\begin{equation*}
\BlendLayer(\ell_1, \Draw(\Empty(), m, p_2), p_1) = \mathsf{clipAll}(\ell_1, m)
\end{equation*}
This is then proven by induction over $\ell_1$,
  using the fact that
  $\ell_1$ contains no \SaveLayer commands,
  its denotation will contain no \BlendLayer{}s.
The base case, where $\ell_1$ is \Empty, is trivial,
  while the inductive case reduces to the following
  in the abstract model:
\begin{gather*}
\dstin\!\left(
\left\{
\begin{array}{ll}
  c_1, & x \in g \\
  \mathsf{Transparent}, & \text{else}
\end{array}
\right\},
\left\{
\begin{array}{ll}
  c_2, & x \in m \\
  \mathsf{Transparent}, & \text{else}
\end{array}
\right\}
\right)\\
=
\left\{
\begin{array}{ll}
  c_1, & x \in g \cap m \\
  \mathsf{Transparent}, & \text{else}
\end{array}
\right\}
\end{gather*}
  where $c_2$ is known to be opaque.
This is proven by case analysis on $x \in g$ and $x \in m$.

\subsection{Subsuming Filters}
\label{subsec:subsume-filters}

\begin{figure}
  \centering
  \begin{minipage}[c]{0.62\linewidth}
\begin{BVerbatim}[fontsize=\scriptsize]
<svg width="350" height="268" viewBox="0 0 350 268">
  <mask id="a" x="0" y="0">
    <path d="M345.05 0H4.95C2.21619 ... 345.05 0Z" fill="white"/>
  </mask>
  <g mask="url(#a)">
    <path d="M345.05 ... 0Z" fill="white"/>
    ...
  </g>
</svg>
\end{BVerbatim}
  \end{minipage}\hfill
  \begin{minipage}[c]{0.25\linewidth}
    \centering
    \frame{\includegraphics[width=\linewidth]{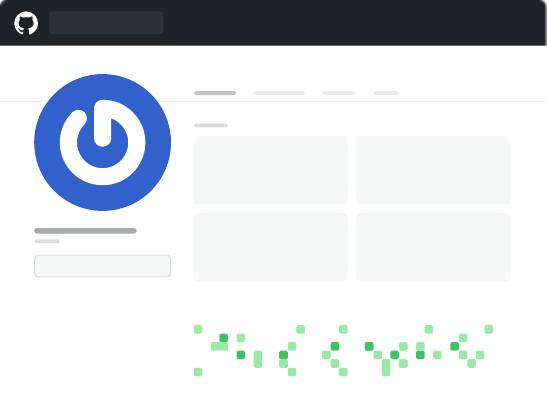}}
  \end{minipage}
  \caption{SVG snippet from \texttt{gravatar.com}, used in this section. This snippet draws the contents of the image, and then masks the contents with a rounded rectangle.}
  \label{fig:subsume-svg}
  \Description{Snippet of SVG code followed by rendering of the same code.}
\end{figure}

\Cref{fig:subsume-svg} shows a snippet of SVG code;
  this snippet specifically is from \texttt{gravatar.com}.
SVG is a declarative, XML-based language for vector graphics
  implemented by Google Chrome.%
\footnote{
It is delightfully recursive that
  SVG is itself a vector graphics language,
  which Chrome executes with Skia;
  and then Skia itself offers vector graphics operations.
}

In this snippet, the \texttt{<g>} element contains two sets of elements.
The first set of elements draws the title bar, logo, and all the various content elements seen in the picture
  using SVG shape constructs like \texttt{path}.
The \texttt{<mask>} element then clips what was drawn with a rounded-rectangle.

Chrome performs the draw-and-mask operation
  by drawing the mask on a \SaveLayer and blending it on the various paths that were drawn
  using \DstIn blending.
However, while \DstIn blending is based on mask layer transparency,
  SVG masking (by default) is based on luminance,
  requiring an extra conversion step.
Chrome performs this conversion using
  a separate \SaveLayer with a luminance color filter.
This is suboptimal:

\begin{center}
  \begin{tabular}{L c L}
    \begin{tabular}{@{}l@{}}
      $\ell\ldots$ \\
      SaveLayer($p$ \{ \DstIn \}); \\
      \quad SaveLayer($p_1$ \{ \textsf{Luma}, \SrcOver \}); \\
      \quad \quad Draw($m$, $p_2$ \{ \textsf{Solid}($c$), \SrcOver, \Id \}); \\
      \quad Restore() \\
      Restore()
    \end{tabular}
    & =
    & \begin{tabular}{@{}l@{}}
      $\ell\ldots$ \\
      SaveLayer($p$); \\
      \quad Draw($m$, $p_3$ \{ Solid(\Luma{}($c$)), \SrcOver, \Id \}); \\
      Restore()
      \end{tabular}
  \end{tabular}
\end{center}

We prove this equivalence unconditionally.
The replacement pattern on the right
  removes the luminance conversion \SaveLayer
  and simply pre-computes the luminance of the mask fill color.
The remaining \SaveLayer in the replacement pattern
  can be removed using the \DstIn-\Clip rewrite
  of \Cref{subsec:dstin-clip}.
However, this equivalence \emph{does not} generalize
  to cases where the inner \SaveLayer contains
  multiple \Draw commands
  due to the way multiple colors can interact
  with the \Luma color filter.

Proving this rewrite in the \muskia semantics
  does not require induction.
Instead, the proof abstracts over
  the state $\Sigma$ produced by $\ell$
  and then evaluates the fixed sequence
  of three Skia commands in the pattern.
This reduces to the following equation in the abstract model:
\begin{equation*}
\Luma(\srcover(\Transparent, c)) = \srcover(\Transparent, \Luma(c))
\end{equation*}
The equation is trivially true
  because $\SrcOver(\Transparent, c) = c$.
An analogous rewrite with two or more \Draw commands,
  by contrast, would require this more general and incorrect equation:
\begin{equation*}
\Luma(\srcover(c_1, c_2)) \ne \srcover(\Luma(c_1), \Luma(c_2))
\end{equation*}

\subsection{Gradient Masks}
\label{subsec:grad-masks}

Three Russian websites---%
  \texttt{mail.ru}, \texttt{dzen.ru}, and \texttt{yandex.ru}%
\footnote{
  Which together received just over 5 billion visits/month in January 2026!
}%
  ---use a common web framework which,
  for some reason,
  uses opaque gradients as masks.
The gradients \emph{do not} actually mask out any content,
  because they are opaque,
  but they are expensive for Skia to raster.
This suggests the following idealized equivalence:
\begin{center}
  \begin{tabular}{L c L}
    \begin{tabular}{@{}l@{}}
      Draw($s$, $p_1$ \{ SrcOver \});\\
      SaveLayer(\{ DstIn \});\\
      \quad Draw($s$, $p_2$ \{ Gradient($g$), SrcOver, \Id \});\\
      Restore();
    \end{tabular}
    & = &
    \begin{tabular}{@{}l@{}}
      Draw($s$, $p_1$); \\
    \end{tabular}
  \end{tabular}
\end{center}
Here the left hand side is
  a complete, four-instruction Skia program;
  this rewrite is performed under \SaveLayer
  on the target web pages.

We prove this equivalence under the condition
  that every color stop in $g$ is opaque.
In this case, the \DstIn blend does not affect the initial \Draw
  and the mask is entirely superfluous.
Again, no induction is needed,
  since only a fixed number of \muskia commands are present.
Converting to \Layer terms, the rewrite unfolds to:

\begin{center}
  \begin{tabular}{L c L}
    \begin{tabular}{@{}l@{}}
      BlendLayer(Draw(Empty(), $s$, $p_1$),\\
      \quad \quad Draw(Empty(), $s$, $p_2$),\\
      \quad \quad \{ DstIn \})
    \end{tabular}
    & = &
    \begin{tabular}{@{}l@{}}
      Draw(Empty(), $s$, $p_1$)
    \end{tabular}
  \end{tabular}
\end{center}

\noindent
This reduces to the following equation in the abstract model:
\[
\DstIn(\SrcOver(\Transparent, a), b) = \SrcOver(\Transparent, a)
\]
\noindent
  where $b$ is known to be opaque
  because all color stops in the gradient $g$ were opaque.

All four suboptimal patterns in this section
  require careful reasoning about colors, blends, and filters
  in order to guarantee that the optimization is correct.
While each high-level rewrite is straightforward,
  we needed many attempts to write down valid replacement patterns
  and discover critical assumptions
  like the lack of \SaveLayer operations in certain command sequences
  or the opacity of the colors involved.
The \muskia semantics, and the rigorous proofs it enables,
  were essential to surfacing assumptions
  and ensuring that each rewrite was correct.

\section{Compiler}
\label{sec:compiler}

To demonstrate the utility of the \muskia semantics
  for developing impactful Skia tooling,
  this section discusses a Skia optimizer
  that applies the rewrites discussed in \Cref{sec:rewrites}.
The optimizer is highly efficient,
  processing thousands of Skia commands in microseconds,
  allowing Skia clients to meet demanding refresh rate targets.

\subsection{Rewriting}

\begin{figure}[h]
  \footnotesize
  \begin{tabular}{@{}cc@{}}
    \begin{minipage}[t]{0.42\linewidth}
      \vspace{0pt}
      \begin{BVerbatim}[baselinestretch=1.1,fontsize=\scriptsize]
template <typename MatchState>
struct SkOptPass {
    struct Frame {
        MatchState match_state;
        int save_count;
    };
    STArray<8, Frame> frames;
    int save_count;

    void transform(SkRecord*);

    MatchState onSaveLayer(SkRecord*, int);
    void onSave(SkRecord*, int);
    void onRestoreLayer(SkRecord*, int);
    void onRestore(SkRecord*, int);
    void onDraw(SkRecord*, int);
    void onClip(SkRecord*, int);
    void onOther(SkRecord*, int);
};
\end{BVerbatim}
    \end{minipage}
    &
    \begin{minipage}[t]{0.54\linewidth}
      \vspace{0pt}
      \centering
      \begin{tikzpicture}[
        scale=0.9,
        transform shape,
        baseline=(current bounding box.north),
        >=to,
        every state/.append style={minimum size=12mm, font=\scriptsize}
      ]
        \node[state] (ignore) at (0,0) {\texttt{Ignore}};
        \node[state] (matching) at (2.6,0) {\texttt{Matching}};
        \node[state, accepting] (rewrite) at (1.3,-2.2517) {\texttt{Rewrite}};
        \draw[->] (0,1.6) -- node[left, font=\tiny]{\SaveLayer(\textit{p}
          \{$\neg$\SrcOver{}\})} (ignore.north);
        \draw[->] (2.6,1.6) -- node[right, font=\tiny]{\SaveLayer(\textit{p}
          \{\SrcOver{}\})} (matching.north);
        \draw[->] (matching.north west) to[bend right=22] node[above, sloped,
        font=\tiny]{\Draw(\textit{s}, \textit{p} \{$\neg$\SrcOver{}\})}
        (ignore.north east);
        \draw[->] (matching) to[bend left=22] node[right,
        font=\tiny]{\Restore{}()} (rewrite);
        \draw[->] (matching) edge[loop right] node[right,
        font=\tiny]{\Draw(\textit{s}, \textit{p} \{\SrcOver{}\})} (matching);
      \end{tikzpicture}
    \end{minipage}
  \end{tabular}
  \caption{Here we show both the definition of the abstract \texttt{SkOptPass}
    rewrite pass,
    and the corresponding state machine implementing the
    \SrcOver-\SaveLayer rewrite. The state machine state is managed by the \texttt{MatchState}
    struct. It tracks the state, and any metadata needed to complete the
    rewrite. In this case the only extra metadata needed is the location of the
    \SaveLayer to be removed. The state machine we show is the state machine for
    only one layer for simplicity. The rewriter actually functions like a stack
    of state machines, with each \SaveLayer pushing the start of a new state
    machine onto a stack, and a \Restore popping it off. Sometimes the actions
    of the top-most state machine can change the state of the ones below. We
    implement this by allowing the rewriter to look beyond the top of the stack.}
  \label{fig:statemachine}
  \Description{The interface of the SkOpt class, followed by the state machine of the SrcOver SaveLayer rewrite.}
\end{figure}

When a Skia client calls a method on the \texttt{SkCanvas} object,
  Skia records this command
  as an \texttt{SkRecords} object
  in an in-memory buffer
  called an \texttt{SkRecord}.%
\footnote{
Yes, the class name for the collection is singular
  while the class name for its elements is plural.
}
The optimizer then searches the \texttt{SkRecord} buffer
  for each rewrite rule from \Cref{sec:rewrites},
  replacing the matched left-hand side commands
  with the right-hand side of the rewrite rule.
Each rewrite is implemented as a single optimization pass
  that performs a single pass over the command buffer.
A single pass can perform multiple matches.

Each optimization pass implements
  the \texttt{SkOptPass} interface shown in \Cref{fig:statemachine}.
Implementation passes subclass from \texttt{SkOptPass}
  and implement the \texttt{on*} callback methods
  to handle different Skia command types.
The \texttt{transform} method is the main entry point,
  performing a single optimization pass
  over an \texttt{SkRecord} command buffer
  by calling the relevant callback method
  for every Skia command in the buffer.
The \texttt{onOther} callback is called
  for all commands outside the \muskia semantics;
  generally the \texttt{onOther} callback
  invalidates the current match.
However, \texttt{SkOptPass} also handles a key gap between
  the actual Skia language
  and the \muskia semantics of \Cref{sec:semantics}:
  matching \Save and \SaveLayer
  to their corresponding \Restore calls
  and thereby calling \texttt{onRestore} or \texttt{onRestoreLayer}.

To do so, the \texttt{SkOptPass} class stores
  a \texttt{save\_count} and a stack of \texttt{Frame} objects.
Every \Save command increments the \texttt{save\_count} field,
  while every \SaveLayer command pushes a new \texttt{Frame},
  which in turn also saves the current \texttt{save\_count}.
When a \Restore command is seen,
  the \texttt{save\_count} field is compared to
  the \texttt{save\_count} of the top \texttt{Frame}
  to determine if a \Save or \SaveLayer is being closed
  and to call the appropriate callback.
Additionally, the \muskia language permits
  rewrites under \SaveLayer commands,
  so in principle a rewrite can apply to any \SaveLayer/\Restore pair.
For this reason, the \texttt{Frame} stack also stores
  a \texttt{MatchState} for each currently-open \SaveLayer;
  the initial \texttt{MatchState} for the layer
  is returned from \texttt{onSaveLayer}.
The stack of frames is stored in a Skia \texttt{STArray},
  a container that stores its first eight entries on the stack.
The stack depth corresponds to the \SaveLayer depth,
  and most Chrome-generated Skia programs we have seen
  have shallow \SaveLayer depths,
  so the frame stack is typically entirely stack-allocated.

The \texttt{MatchState} type is instantiated by
  every individual optimization pass
  and implements a finite state automaton
  that matches the rewrite rule.
For example, the \SrcOver-\SaveLayer rewrite
  uses two states, \texttt{Matching} and \texttt{Ignore},
  shown in \Cref{fig:statemachine}.
The initial \texttt{MatchState} is \texttt{Matching}
  for \SrcOver-\SaveLayer commands%
\footnote{
The optimizer also requires
  the \SaveLayer command to be opaque;
  while \Cref{sec:semantics} does not include opacity,
  our Lean formalization does.
}
  and \texttt{Ignore} otherwise.
The \texttt{MatchState} also stores
  the index in the \texttt{SkRecord}
  of the \SaveLayer command,
  for later rewriting.
\SrcOver-\Draw commands stay in the \texttt{Matching} state,
  while all other \Draw commands and all \SaveLayer commands
  transition to the \texttt{Ignore} state.
All other commands (\Clip, \Save, \Restore)
  do not affect the match state.
The \texttt{MatchState} therefore indicates
  whether a given layer
  is a \SrcOver-\SaveLayer containing only
  \SrcOver-\Draw commands---%
  exactly the requirements of the \SrcOver-\SaveLayer rewrite
  in \Cref{sec:rewrites}.
The other rewrites in \Cref{sec:rewrites} have
  similar, though more elaborate, state machines.
In \texttt{onRestoreLayer},
  if the current state is matching,
  the actual rewriting is performed
  by changing the \SaveLayer command
  into a \Save command
  using the saved index.

\subsection{Memory Management}

While the single-pass design does limit
  the computational complexity of our optimizer,
  efficiently manipulating Skia programs in memory
  is also critical to hitting demanding latency limits.
Each \texttt{SkRecords} object
  has a tag and fields for all the arguments
  to that specific command.
\texttt{SkRecords::DrawRect}, for example,
  has a \texttt{DrawRect\_Type} tag,
  an \texttt{SkRect} for the rectangle bounds,
  and an \texttt{SkPaint} for the paint parameters.
The \texttt{SkRecord} buffer stores two fields:
  \texttt{fRecords},
  an array of pointers to \texttt{SkRecords} objects,
  and \texttt{fAlloc}, an arena allocator
  where the \texttt{SkRecords} objects are stored.
 \Cref{fig:skrecord-layout} illustrates the memory layout.
Importantly, this design means that
  the compiler can mutate the \texttt{SkRecord} buffer
  without moving the actual \texttt{SkRecords} objects,
  simply by adjusting the pointers in \texttt{fRecords}.

\begin{figure}[tbp]
  \centering
  \begin{tikzpicture}[
    x=1cm,
    y=1cm,
    >={Latex[length=2mm]},
    ptr/.style={draw, minimum width=0.8cm, minimum height=0.6cm, inner sep=0pt, font=\scriptsize, align=center},
    rec/.style={draw, minimum width=2.9cm, minimum height=0.8cm, font=\scriptsize, align=center},
    lbl/.style={font=\small\ttfamily}
  ]
    \node[lbl, anchor=east] at (-0.3, 1.2) {fRecords:};
    \node[lbl, anchor=east] at (-0.3, 0.0) {fAlloc:};

    \node[ptr] (p1) at (0.05, 1.2) {$*$};
    \node[ptr, right=0cm of p1] (p2) {$*$};
    \node[ptr, right=0cm of p2] (p3) {$*$};
    \node[anchor=west, font=\small] at ($(p3.east)+(0.10,0)$) {$\cdots$};

    \node[rec] (r1) at (1.1, 0.0) {\texttt{SkRecords::DrawRect}};
    \node[rec, right=0cm of r1] (r2) {\texttt{SkRecords::ClipRect}};
    \node[rec, right=0cm of r2] (r3) {\texttt{SkRecords::DrawRect}};
    \node[anchor=west, font=\small] at ($(r3.east)+(0.10,0)$) {$\cdots$};

    \draw[->, shorten >=2pt] (p1.south) -- (r1.north west);
    \draw[->, shorten >=2pt] (p2.south) -- (r2.north west);
    \draw[->, shorten >=2pt] (p3.south) -- (r3.north west);

  \end{tikzpicture}
  \caption{
    \texttt{SkRecord} stores an array of pointers (\texttt{fRecords})
    into a contiguous chunk of record objects (\texttt{fAlloc}),
    each of which represents a single Skia command.
  }
  \Description{A block diagram of the memory layout of SkRecord}
  \label{fig:skrecord-layout}
\end{figure}

Applying the rewrites from \Cref{sec:rewrites}
  requires changing, removing, or adding commands
  to the \texttt{SkRecord}.
Changing commands is easy:
  the optimizer allocates a new \texttt{SkRecords} object in the arena
  and changes the \texttt{fRecords} buffer to point to this new object.
Removal likewise changes a command
  into a special \texttt{SkRecords::NoOp} command.
When inserting new commands, however,
  it would be expensive to move the \texttt{fRecords} elements,
  especially since multiple rewrites can apply to the same \texttt{SkRecord}.
We therefore store pending insertions into
  a separate insertion buffer,
  and merge the original and insertion buffer
  in a single linear scan
  after the optimization pass is complete.
We learned of this technique from the WebKit browser engine's B3 JIT,
  where it is called an ``InsertionSet''~\cite{insertionset},
  though the general technique is well-known.

Skia can serialize \texttt{SkRecord} objects
  to and from a binary format called a ``Skia Picture'' or \texttt{skp}.
All information in a \texttt{SkRecord} format required for drawing,
  including referenced external images,
  are serialized into the \texttt{skp}.%
\footnote{
Though typefaces do not get serialized into the \texttt{skp}.
}
Google Chrome can dump generated Skia programs as \texttt{skp}s
  with the \texttt{chrome.gpuBenchmarking.printToSkPicture} debug method,
  available when Chrome is run with \texttt{--no-sandbox} and
  \texttt{--enable-gpu-benchmarking} flags.
It is important to note here that Chrome generates
  multiple Skia programs, meaning multiple \texttt{skp}s,
  for a single website.
These multiple programs represent different ``layers'' of the page,
  which are then composited in a later step;
  this split into layers is chosen by Chrome
  to optimize animations~\cite{wbe}
  and is unrelated to Skia's concept of layers.
Our optimizer operates on
  individual \texttt{SkRecord} objects in memory
  for maximum performance,
  but our evaluation does save and load
  these \texttt{SkRecord} objects from/to \texttt{skp} files
  to minimize interference from Google Chrome during benchmarking,
  as discussed in more detail in \Cref{sec:eval}.

\section{Evaluation}
\label{sec:eval}

We evaluate the \muskia semantics, and the optimizer based on them,
  on a collection of real-world Skia programs
  generated by Google Chrome on the Top 100 websites,
  focusing on four research questions:

\begin{enumerate}
\item[RQ1] Is the optimizer correct? 
\item[RQ2] Is the optimizer effective? 
\item[RQ3] Is the optimizer fast? 
\item[RQ4] Are the optimizations generic across Skia backends, hardware, and websites? 
\end{enumerate}

\subsection{Methodology}

We evaluate our optimizer on Skia programs
  generated by Google Chrome on the
  top 100 most-visited websites in the world,%
\footnote{Excluding a few pornographic websites.}
  using the list collected by \texttt{tranco-list.eu}~\cite{trancolist}.%
\footnote{The famous ``Alexa Top 100'' is no longer produced.}
The list collected from \texttt{tranco-list} also includes
  infrastructural domains like \texttt{gltd-servers.net}
  which do not have any user facing websites, making them unusable for our evaluation.
So we further filter this list through the Chrome User Experience Report (CrUX)~\cite{crux}
  which explicitly collects user-facing websites (but does not rank websites based on visits).
Google Chrome, via the \texttt{playwright} Python library, is used to visit
  the main page of each website,
  and \texttt{skp} files are dumped as described in \Cref{sec:compiler}.
Each website produces one \emph{or more} \texttt{skp} files.
Since our optimizer focuses on reducing
  the number of \texttt{SaveLayer} commands,
  we focus on \texttt{skp} files that have \texttt{SaveLayer} commands.
For our primary evaluation in RQs 1--3,
  this leaves \graphiteAppleNumBench benchmarks across \N websites.

The optimizer itself is based off of Skia commit \texttt{f931a2e},
  on the \texttt{main} branch between milestones 143 and 144~\cite{skia-releases}.
For our primary evaluation in RQs 1--3,
  we build Skia with Clang version 17.0.0 with \texttt{-O3} and \texttt{-flto=thin}
  and run it on an 8-core Apple M2 unified CPU/GPU with 16GB RAM
  using Skia's ``Graphite'' graphics backend and the Metal GPU API.

\subsection{RQ1: Is the Optimizer Correct?}
\label{subsec:rq1}

This section aims to answer RQ1 by showing
  that the optimizer is correct,
  in two senses: that it conforms with the \muskia semantics
  and that it also does not introduce visible distortions
  on the evaluation websites.

\subsubsection{Conformance with \muskia Semantics}

To test whether the optimizer conforms with the \muskia semantics,
  we extended it with a mode that outputs a \texttt{skp}
  after each individual rewrite pass,
  forming an ``optimization trace''.
Each of these \texttt{skp} files is then converted
  into a readable JSON format using \texttt{skp\_parser},
  a tool distributed with Skia,
  and then read and converted to \muskia Lean terms
  using a short Python program.
Not every \texttt{skp} can be converted,
  since \muskia does not formalize some advanced Skia features
  like mask and image filters,
  but for \tvCompilePassN benchmarks
  all steps in the optimization trace convert cleanly.
The conversion program also performs some minor simplifications,
  such as down-casting \texttt{SkPath} shapes
  to simpler shapes like \texttt{SkRect} when possible,
  to invert a transformation done by Skia.
If all \texttt{skp} files in the optimization trace
  can be converted to a \muskia Lean term,
  a sequence of lemmas is asserted in Lean,
  claiming that each step in the trace
  is equivalent, as a \muskia program, to the previous step.

\begin{figure}
\small
\begin{BVerbatim}
theorem src_eq_opt1 : denote src = denote opt1 := by
  unfold src
  unfold opt1
  grind (gen := 80) [GradientMask]
\end{BVerbatim}
  \caption{
  Example of a translation validation lemma.
    \texttt{src} is the source \muskia term,
    and \texttt{opt1} is the term after one optimization pass.
  The relevant rewrite theorem, \texttt{GradientMask},
    is passed to \texttt{grind} to search for a proof;
    a number of general-purpose \muskia theorems
    are also included by \texttt{@[grind]} annotations.
  The \texttt{gen} argument to \texttt{grind}
    increases its search fuel
    to enable proofs for larger \muskia terms.}
  \Description{Theorem statement and proof used by translation validation.}
\label{fig:layerTV}
\end{figure}

To prove this theorem, we rely on Lean's \texttt{grind} tactic.
The \texttt{grind} database is loaded with
  theorems for the rewrite rules applied by the optimizer in this step
  as well as general theorems about \muskia equivalence.
\Cref{fig:layerTV} shows an example proof.
Translation validation succeeds for \tvPassN{}
  out of \graphiteAppleNumBench benchmarks,
  demonstrating that our optimizer conforms with the \muskia semantics
  and providing a ``foundational'' certification of its correctness
  on these benchmarks.
Some of the successful translation validations are quite complex;
  for example, the main \texttt{skp} from Pinterest contains
  a luminance filter nested in a \DstIn mask nested in an opaque \SaveLayer.
Three separate optimization passes fire on this benchmark,
  with each one exposing the next rewrite opportunity;
  all three translation validate.
The remaining \tvGrindN benchmarks
  do not translation-validate due to \texttt{grind} timeouts;
  as a general-purpose proof search engine,
  \texttt{grind} is slower than our optimizer
  and times out on complex \muskia terms.

\subsubsection{Optimizations Yield the Same Images}

Conformance with the \muskia semantics, however,
  is only as good as the quality of the \muskia semantics themselves.
We therefore compare the baseline and optimized \texttt{skp}
  by rasterizing both \texttt{skp} to PNG images
  and performing a pixel diff using ImageMagick \texttt{compare}.
The \muskia rewrites of \Cref{sec:rewrites}
  are proven in a real-valued abstract model,
  so ignore effects from discretization and anti-aliasing,
  which are caused by the discrete pixel grid and finite color depth,
  so some differing pixels are expected;
  we therefore pass \texttt{-metric AE} and \texttt{-fuzz 1\%}
  to \texttt{compare},
  asking it to report the number of pixels
  that differ by more than \texttt{1\%} in a color channel.
Notably, we perform the pixel comparison on all benchmarks,
  and can thus test the optimizer's correctness
  even for benchmarks that use features
  not formalized in \muskia.

\begin{figure}
  \includegraphics[scale=0.7]{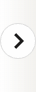}\hspace{0.5em}
  \includegraphics[scale=0.7]{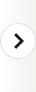}\hspace{0.5em}
  \includegraphics[scale=0.7]{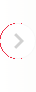}
  \caption{
    On the left, a button drawn by the \texttt{skp}
      captured for Layer 45 on \texttt{yahoo.com}.
    In the middle,
      the same button drawn by the optimized \texttt{skp}.
    On the right, the same button is drawn, lighter,
      with differing pixels drawn in red.
    The differing pixels are clearly placed along
      a curved shape edge,
      exactly where anti-aliasing effects are expected.
  }
  \Description{Figure showing 3 images side-by-side, the left is the unoptimized skp, the middle is the optimized, and the right is the diff.}
\label{fig:diff}
\end{figure}

Across the \graphiteAppleNumBench benchmarks,
  only \graphiteAppleNumWithPixDiff report significant pixel differences.
We examine all such instances manually;
  \Cref{fig:diff} shows one example, from Yahoo.
In all such instances, the differing pixels
  are clearly caused by anti-aliasing,
  as evident by the ``spotted'' nature of differences
  and their location along curved shape edges.
The authors could not tell
  the baseline and optimized rasterizations apart by sight.
Skia does not guarantee its anti-aliasing behavior,
  which can further differ by monitor type or operating system,
  and we confirmed with a member of the Skia team
  that anti-aliasing changes are acceptable.

\subsection{RQ2: Is the Optimizer Effective?}
\label{subsec:rq2}

This section aims to answer RQ2 by showing
  that the optimizer is effective
  at reducing the rasterization time of Skia programs.

To measure raster time, we use \texttt{nanobench},
  Skia's standard, internal benchmarking tool.
We specifically use the \texttt{playback} benchmarking option,
  which rasters the same exact \texttt{skp} in a loop,
  calibrating the number of loops so that
  the total GPU rasterization time is \SI{5}{ms}.
This makes timing overhead negligible.
\texttt{nanobench} also performs warm-up iterations
  to avoid measuring shader compilation time.
Each call to \texttt{nanobench} measures the rasterization time
  for each \texttt{skp} 100 times,
  and we use the geometric mean of these samples.

However, during the development and testing of the optimizer,
  we found that individual calls to \texttt{nanobench} were still noisy.
This variability may result from GPU load,
  scheduling,
  other system effects,
  or the \texttt{skp} serialization and deserialization machinery.
Hence, we perform 25 independent \texttt{nanobench} runs for each benchmark,
  compute a speedup from the geomean runtime of each invocation,
  and report the median of these speedups
  together with a 95\% bootstrap confidence interval.

\begin{figure}[t]
  \centering
  \begin{minipage}[t]{0.49\linewidth}
    \centering
    \includegraphics[scale=0.5]{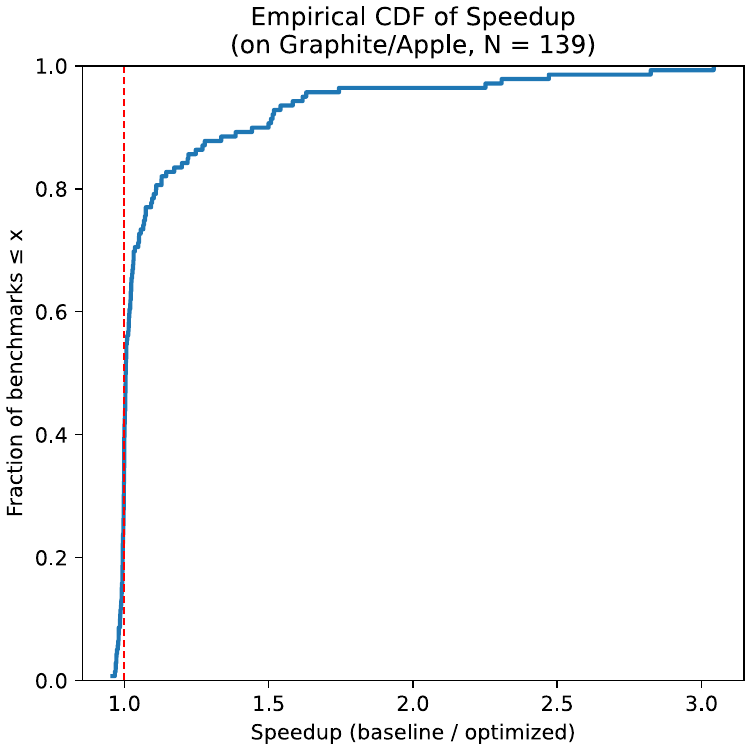}
    \captionof{figure}{
      Raster-time speed-up CDF.
      The horizontal axis shows a speedup value,
      with the red line marking a 1\texttimes{} speedup
      meaning identical runtime for the baseline and optimized \texttt{skp}.
      The vertical axis shows the fraction of benchmarks
      that meet or exceed that speed-up amount.
      We find a geo-mean speedup of \graphiteAppleGeoSpeedup across the benchmarks.
      }
    \label{fig:grmetal-cdf}
  \end{minipage}\hfill
  \begin{minipage}[t]{0.49\linewidth}
    \centering
    \includegraphics[scale=0.5]{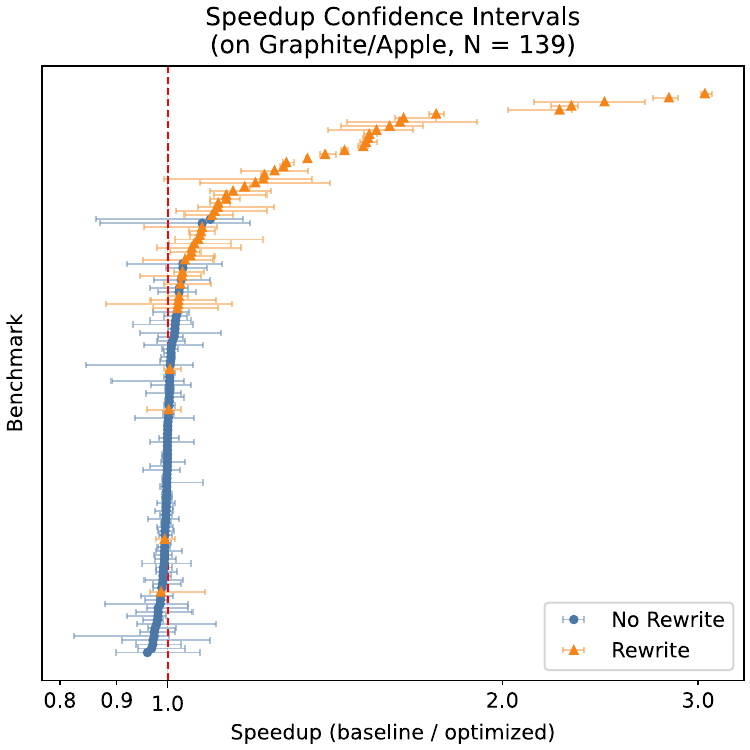}
    \captionof{figure}{
      95\% Confidence Intervals of the geo-mean speedup
        from 25 \texttt{nanobench} runs for all \graphiteAppleNumBench benchmarks.
      The \orange{orange} intervals are from benchmarks that have been rewritten
        by the optimizer.
      The \blue{blue} intervals are from benchmarks that have not been rewritten
        by the optimizer.
      We consider a speedup for a benchmark to be genuine and not influenced by noise
        when the entire interval is to the right of the 1\texttimes\xspace line.
      }
    \label{fig:grmetal-ci}
  \end{minipage}
  \Description{CDF and confidence intervals for Graphite/Apple}
\end{figure}

\Cref{fig:grmetal-cdf} shows our results.
In this CDF, benchmarks are arranged vertically,
  and the dotted red line corresponds to ``no speedup''.
Points to the right of the line get faster when optimized,
  with a maximum speedup of \graphiteAppleMaxSpeedup.
The average geo-mean speedup
  across all \graphiteAppleNumBench benchmarks
  is \graphiteAppleGeoSpeedup.
The very largest speedups are seen
  on layers from \texttt{vimeo.com} (3.04\texttimes) which uses 3 rewrites in succession:
  subsuming a luminance filter,
  converting a \DstIn mask into a \Clip,
  and removing an opaque \SaveLayer.
Each succesive rewrite is enabled by the previous one.
We also see very large speedups on \texttt{dzen.ru} (2.41\texttimes),
  which only uses the gradient mask rewrite
  that uses the framework discussed in \Cref{subsec:grad-masks}.
We also draw all the confidence intervals of each benchmark in \Cref{fig:grmetal-ci}.
38 of the 51 rewritten benchmarks have intervals entirely above 1\texttimes,
  suggesting that their measured speedups are unlikely to be explained by run-to-run noise alone.
The remaining 13 rewritten benchmarks have confidence intervals that cross 1\texttimes{} speedup,
  but none have an interval entirely below 1\texttimes, indicating that the optimizer
  does not significantly slow down any rewritten benchmark.
In short, the optimizer is effective
  at significantly improving the rasterization time
  of real-world Skia programs.

\subsection{RQ3: Is the Optimizer Fast?}

This section aims to answer RQ3 by showing
  that the optimizer is fast enough
  to ``pay for itself'' in rasterization time reductions.

\begin{figure}[t]
  \centering
  \begin{minipage}[t]{0.49\linewidth}
    \centering
    \includegraphics[scale=0.5]{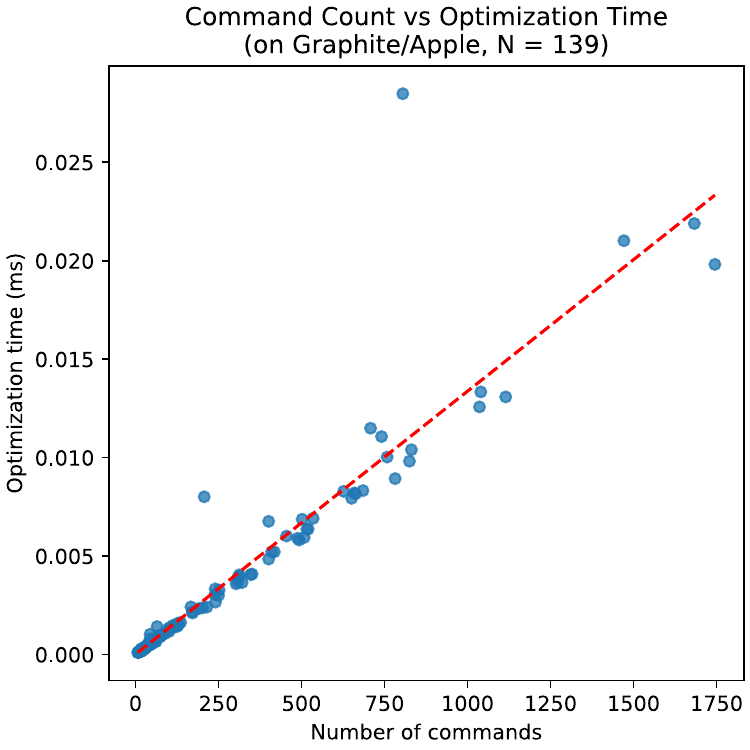}
    \captionof{figure}{Optimization time versus \texttt{skp} size on
      Graphite/Metal. Runtime scales approximately linearly, and the all
      benchmarks optimizes in under \graphiteAppleMaxOptTime}
    \label{fig:grmetal-optscatter}
  \end{minipage}\hfill
  \begin{minipage}[t]{0.49\linewidth}
    \centering
    \includegraphics[scale=0.5]{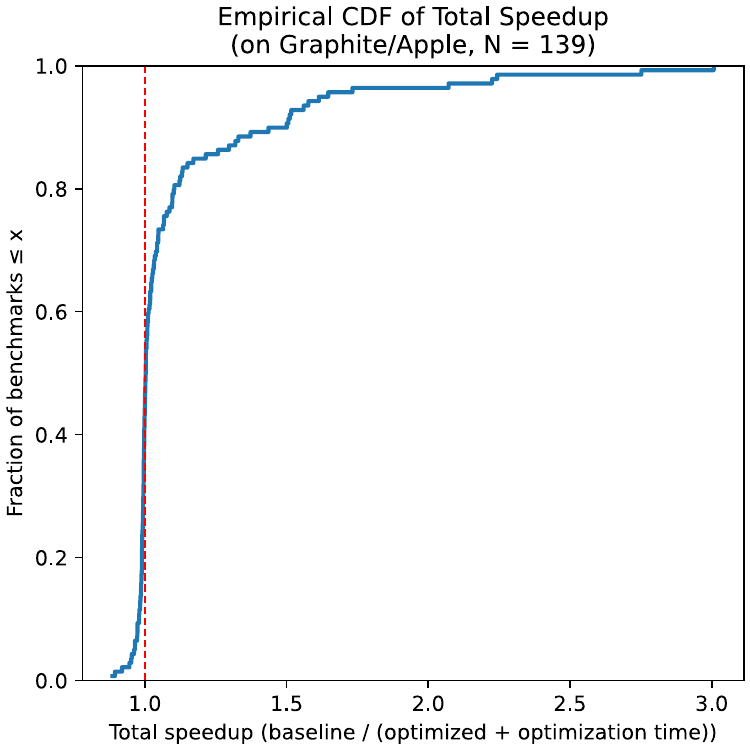}
    \captionof{figure}{CDF with optimization time included.}
    \label{fig:grmetal-optcdf}
  \end{minipage}
  \Description{Scatter plot for optimization time and total time CDF.}
\end{figure}

Skia rasterization must be extremely fast:
  browser engines need rasterization to take
  hundreds of microseconds for even large, complex websites
  to meet their frame budgets.
We thus measure the total optimization time
  using a similar measurement scheme as in RQ2.
\Cref{fig:grmetal-optscatter} shows the results.
On the left, the scatter plot shows that
  the optimizer's runtime is linear in the \texttt{skp} size,
  and all \texttt{skp}s optimize under \graphiteAppleMaxOptTime.
The vast majority of \texttt{skp}s are much smaller
  and thus much faster.
On the right, the CDF of \Cref{fig:grmetal-cdf} is redrawn,
  but with the estimated optimization time
  added to the estimated rasterization time.
Slowdowns are now somewhat more common,
  since some \texttt{skp}s see no optimizations applied
  but still spend time in the optimizer.
Still, the majority of \texttt{skp} continue to show speed-ups,
  with the largest speed-ups still as large as \graphiteAppleMaxTotalSpeedup.

\subsection{RQ4: Generalizability}
\label{subsec:rq4}

This section attempts to answer RQ4
  by showing that the optimizer is effective
  on a range of back-ends, hardware, and websites.
We focus on rasterization time,
  since optimization does not invoke the back-end
  and thus cannot vary with the chosen back-end.

\subsubsection{Backend Generalizability}

\begin{figure}
  \centering
  \includegraphics[scale=0.5]{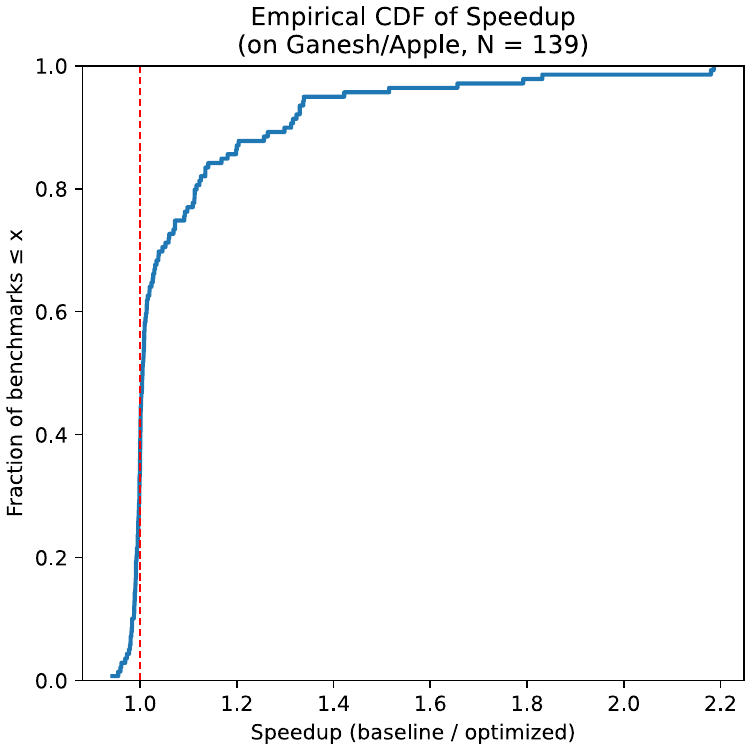}
  \hspace{3em}
  \includegraphics[scale=0.5]{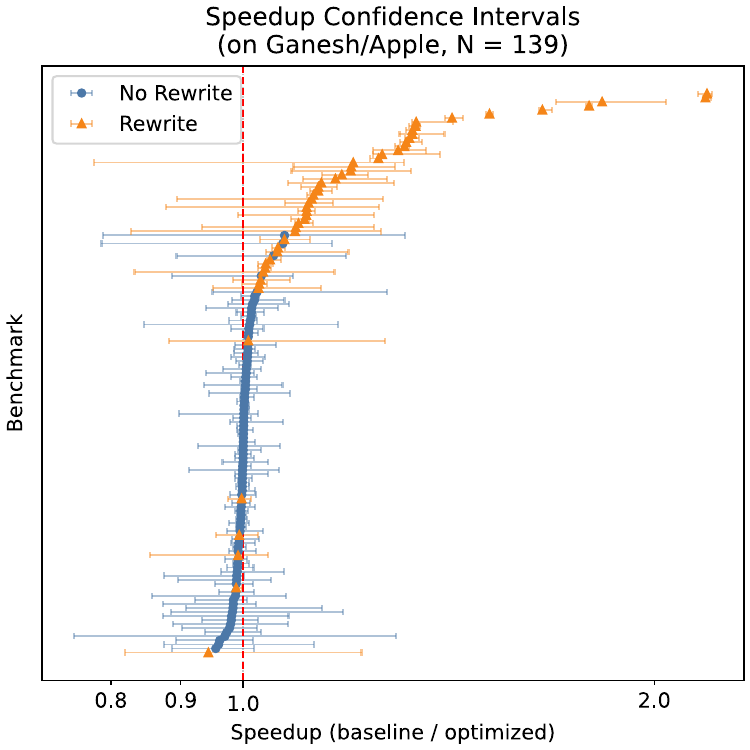}
  \caption{
  Raster-time CDFs and 95\% confidence intervals for the
    Ganesh/Apple (right) backend.
  The newer Graphite backend achieves
    larger speed-ups and fewer slow-downs,
    but the optimizer is still effective on both.
  Graphite/Metal sees a geo-mean speedup of \graphiteAppleGeoSpeedup
    and Ganesh/Apple sees a geo-mean speedup of \ganeshAppleGeoSpeedup.
  }
  \label{fig:cdfs-backend}
  \Description{CDF and confidence intervals for Ganes/Apple}
\end{figure}

To test different Skia backends,
  we compare the speedups achieved by the optimizer
  on the standard Graphite backend,
  which uses the Metal GPU API,
  and on the legacy Ganesh backend,
  which uses the older OpenGL API.
\Cref{fig:cdfs-backend} shows the results for the Ganesh backend.
The optimizer is effective with both contexts,
  though the newer Graphite backend
  does see more substantial speed-ups
  as well as fewer slow-downs.
We suspect that the larger speed-ups result
  from the generally-better performance of Graphite.
\Cref{fig:cdfs-backend} also shows the 95\% confidence intervals for all benchmarks
  running on the Ganesh backend.
Like in Graphite, the majority of the rewritten benchmarks have CI's squarely above 1\texttimes{}
  and none of the rewritten benchmarks are below 1\texttimes{}.

We were also able to reproduce
  Graphite's claimed speedup
  (we measured a geo-mean of 37\%)
  over Ganesh on fixed hardware.

\subsubsection{Hardware Generalizability}

\begin{figure}
  \centering
  \includegraphics[scale=0.5]{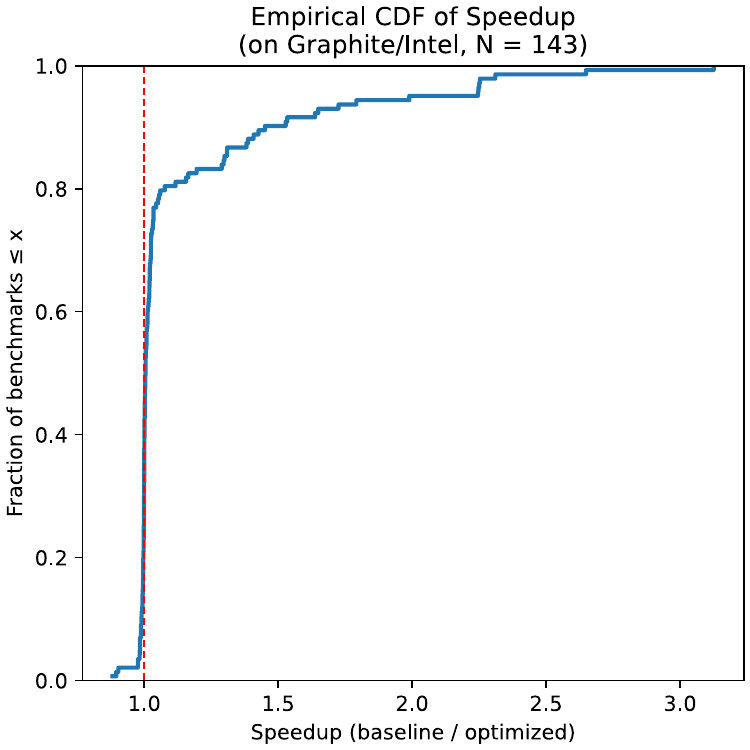}
  \hspace{3em}
  \includegraphics[scale=0.5]{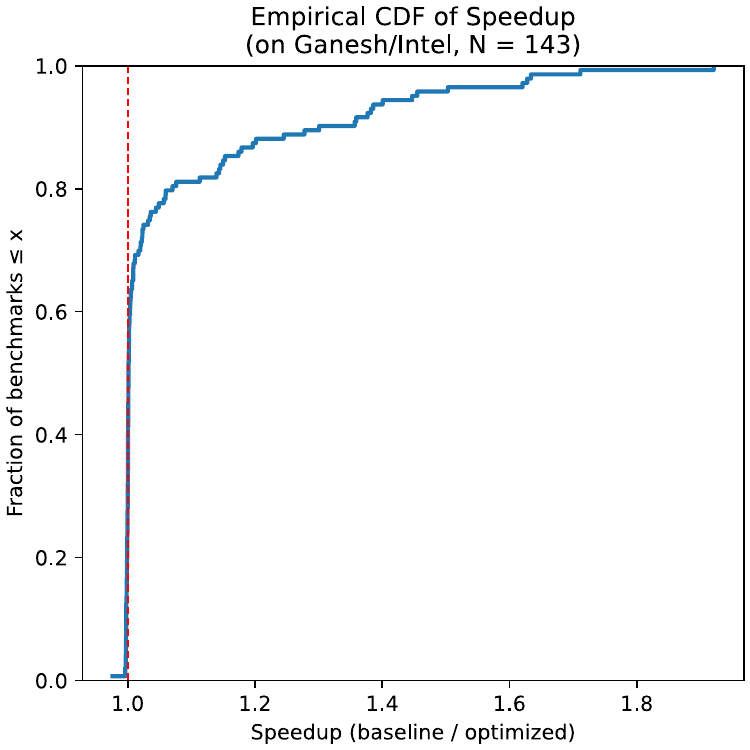}
  \caption{
    Raster-time CDFs for the Graphite/Intel (left) and
      Ganesh/Intel (right) backends on Machine \#2.
    Graphite/Intel sees a geo-mean speedup of \graphiteIntelGeoSpeedup
    and Ganesh/Intel sees a geo-mean speedup of \ganeshIntelGeoSpeedup.
  }
  \label{fig:cdfs-hardware}
  \Description{CDF for Graphite/Intel and Ganesh/Intel}
\end{figure}

\begin{figure}
  \centering
  \includegraphics[scale=0.5]{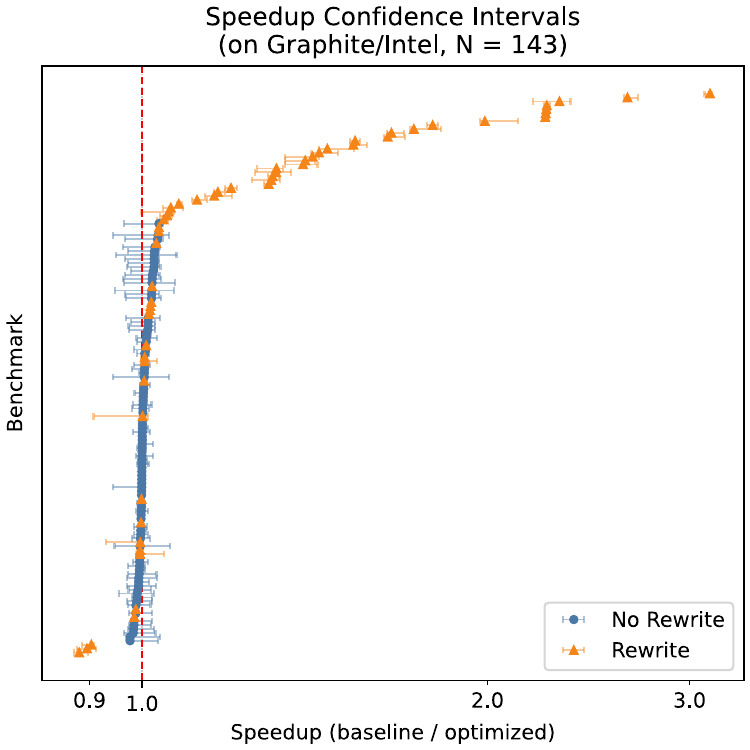}
  \hspace{3em}
  \includegraphics[scale=0.5]{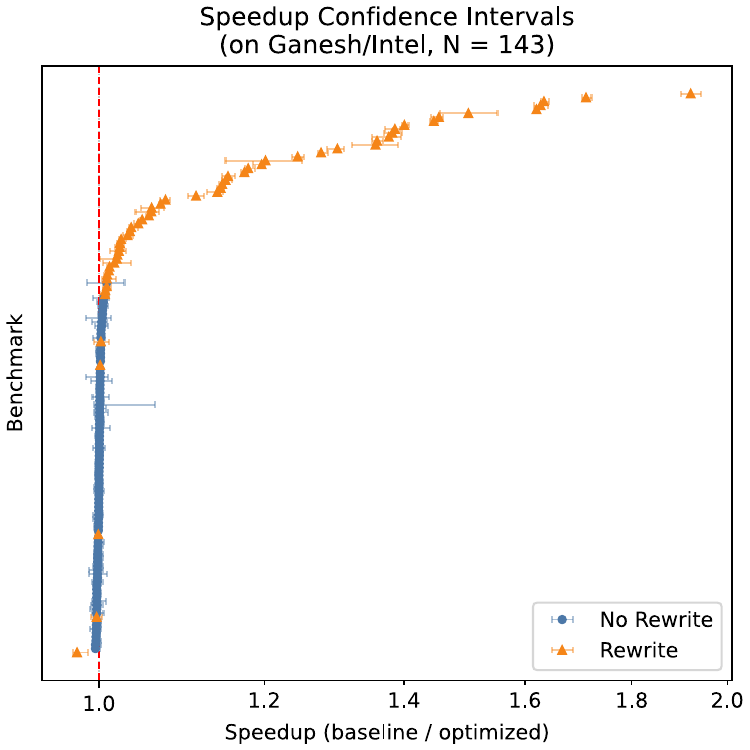}
  \caption{
    95\% confidence intervals for Graphite/Intel (right) and Ganesh/Intel (left).
    Graphite/Intel sees 3 statistical slowdowns and Ganesh/Intel sees only 1.
  }
  \label{fig:cis-hardware}
  \Description{Confidence intervals for Graphite/Intel and Ganesh/Intel}
\end{figure}

To test the effect of different hardware,
  we repeated our evaluation on a second machine,
  which runs Ubuntu 24.04.4 on
  an Intel i7-8700K CPU and an Intel CoffeeLake-S GT2 GPU.
This machine uses Clang version 18.1.3
  and supports the Vulkan and OpenGL graphics APIs.
Chrome's generated Skia programs differ
  depending on the hardware used
\footnote{
Chrome will split the same web page
  into more or fewer layers depending on
  the available hardware.
It will also use different fonts.
Plus, the evaluations were performed
  at different times and may capture
  different versions of the websites in question.
}
  so the benchmark set is re-captured on this machine,
  resulting in \graphiteIntelNumBench benchmarks.
The resulting \texttt{skp}s are optimized
  and measured with both the Graphite
  and Ganesh backends.

\Cref{fig:cdfs-hardware} shows the results.
The optimizer is effective on the second machine too.
The maximum speed-up with the Graphite/Intel backend
  is somewhat larger but still in line with the numbers we see on the Graphite/Apple backend
  (\graphiteIntelMaxSpeedupMatched instead of \graphiteAppleMaxSpeedupMatched).
Looking at the 95\% confidence intervals for the Graphite backend in \Cref{fig:cis-hardware},
  majority of the rewritten benchmarks have CI's squarely above 1\texttimes.
Three benchmarks have CI's below 1\texttimes.
The first regression is from \texttt{intuit.com} layer \#15.
This layer just has one opaque \SrcOver-\SaveLayer removed.
Of note, the first \Draw command in this \SaveLayer is a solid-white fill.
After inspecting the GPU traces for the rasterization of the baseline and optimized versions of this \texttt{skp},
  we see that in the baseline, Graphite seems to be able to optimize away the solid-white fill,
  as it is the first command in a \Layer.
But once the \SaveLayer is removed, this can no longer be done,
  and Graphite allocates a GPU pipeline to do the solid fill.
However, because this \texttt{skp} contains only 25 commands,
  even this small additional cost becomes significant relative to the total rasterization time.
The other two regressions are from \texttt{gravatar.com},
  whose measured slowdowns we found to be rather inconsistent.

With the Ganesh backend,
  the same pattern persists
  (with an even lower maximum speed-up of \ganeshIntelMaxSpeedupMatched),
  but we do find 1 benchmark with a CI under 1\texttimes: \texttt{avast.com} Layer \#0.
This benchmark mainly triggers the \DstIn to Mask rewrite described in detail in \Cref{subsec:dstin-clip}.
But unlike the example used to explain the rewrite,
  which uses a rounded rectangle as its mask,
  this benchmark uses a very complicated path.
Applying this rewrite converts the path into a
  \texttt{clipPath} command with a complex clipping path.
We suspect that the complex clipping path
  somehow triggers bad behavior
  in the older Ganesh backend.
If support for Ganesh were important,
  the \DstIn mask optimization pass
  could be disabled for complex paths.
In any case, this benchmark is
  the only benchmark significantly slowed down by the optimizer on this backend,
  suggesting that the optimizer generalizes well
  across different hardware platforms.

\subsubsection{Website Generalizability}

\begin{figure}
  \centering
  \includegraphics[scale=0.5]{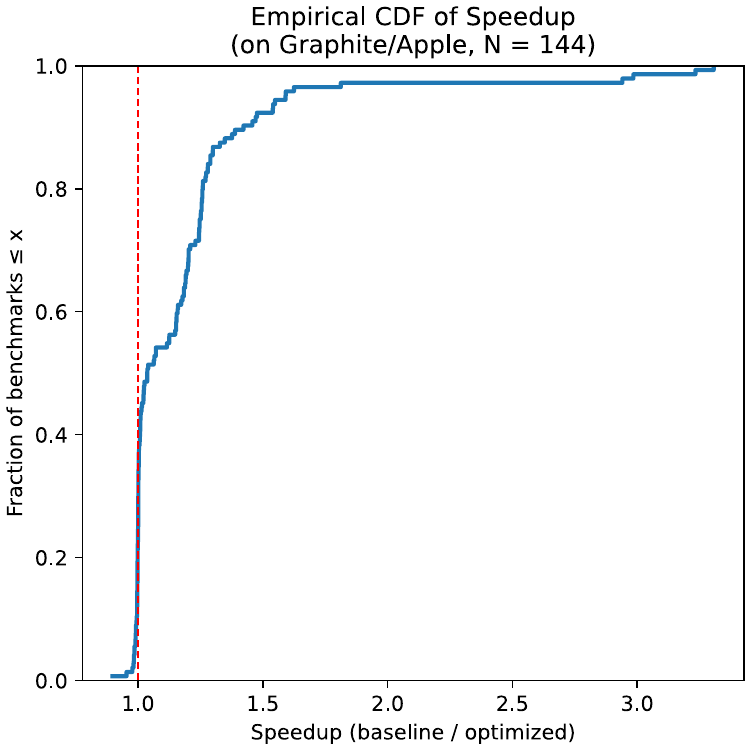}
  \hspace{3em}
  \includegraphics[scale=0.5]{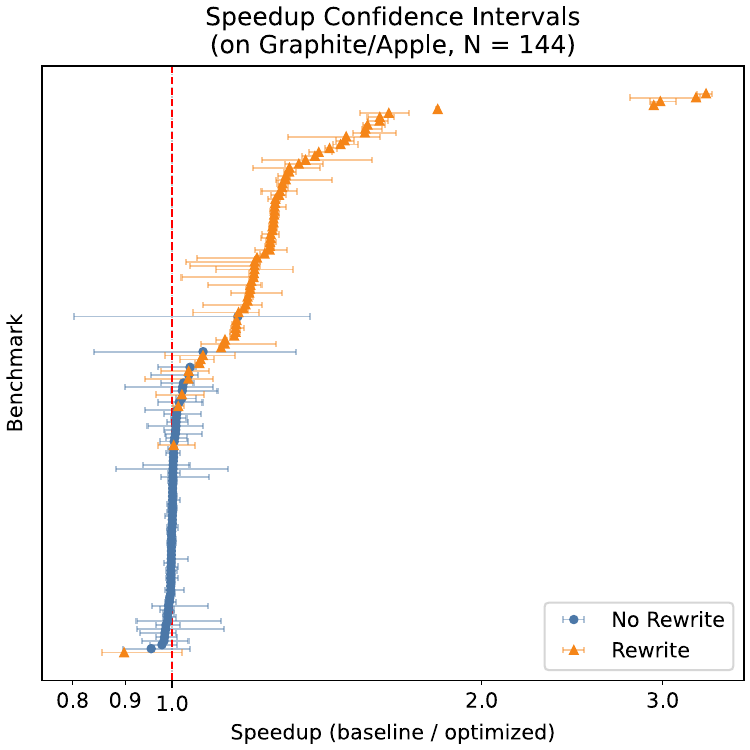}
  \caption{
    Raster-time CDF and 95\% confidence intervals of Graphite/Metal with the new set of 100 websites.
  }
  \label{fig:cdf-newsites}
  \Description{CDF and confidence intervals for Graphite/Metal on next 100 top websites}
\end{figure}

Finally, we ask whether the optimizer is effective
  on different websites than the ones the authors examined.
To test this, we froze the optimization passes and their order
  and performed the same \texttt{skp} gathering process
  for the next 100 most-visited websites,
  which yielded \nextGraphiteAppleNumBench new benchmarks.
The authors have never examined
  the generated Skia programs for these websites
  and have made no attempt to
  tune the optimizer to perform better on these benchmarks.

\Cref{fig:cdf-newsites} shows the results.
Like in the old benchmarks set,
  we see significant speedups of up to \nextGraphiteAppleMaxSpeedup,
  and, like before,
  websites actually rewritten by the optimizer
  show significantly larger speed-ups than
  those that see slowdowns.
All of the slowdowns are websites unaffected by the optimizer,
  so the slowdowns are just noise.
Considering the optimization passes individually,
  the gradient mask optimization fires only on 2 \texttt{skp}s,
  both from \texttt{ivi.ru}.
This validates our approach for developing a rewrite tailored for one popular framework,
  and we expect that analyzing \texttt{skp}s from many websites that use other popular frameworks,
  like Tailwind,
  will also yield good rewrites.
Moreover, the other, more general, optimization passes all see use
  on the new set of benchmarks.

We expect that this new set of benchmarks
  has new, suboptimal patterns
  in its Chrome-generated Skia programs,
  and that other Skia users have their own
  patterns of suboptimal Skia programs.
We expect the \muskia semantics and Lean formalization,
  plus the compiler architecture detailed in \Cref{sec:compiler},
  to prove useful in writing more optimal replacement patterns,
  proving them correct, and enabling our optimizer
  to speed up rasterization for those clients as well.

\section{Discussion and Future Work}
\label{sec:discussion}

We initially began this project by attempting to develop the optimizer directly,
  but discovered that it was far too difficult to write correct rewrite rules:
  as we have discussed in the paper, most require various side conditions,
  and developing the side conditions requires careful reasoning
  about both geometry and the semantics of Skia,
  which at the time we didn’t have a good grasp of.
An example of this is the complexity involved in dealing with \Clip{s}.
The \Clip command,
  together with the \Transform command described in \Cref{subsec:lean},
  are imperative in nature:
  they modify, instead of replacing, the clip (and transform) state.
So, we were ``forced'' to make the \Layer language to make this clip (and transform) state explicit and functional.
We thus changed our focus to develop the formalization,
  which then enabled us to rapidly develop the rewrite rules (described in \Cref{sec:rewrites}) and
  the optimizer (described in \Cref{sec:compiler}).

Our experience suggests several directions for future work on improving
  the formalization and optimization of Skia.

Skia supports many representations of shapes,
  including rectangles, pixel-boundary-aligned rectangles,
  rounded rectangles, polygons, paths, and text.
Although paths provide a general representation, Skia often handles more specific shape types through specialized code paths.
Transformations between these representations can also be confusing;
for example, the intersection of two rectangles is a rectangle,
  but the intersection of two rounded rectangles is a path.
Currently, our abstract model coalesces all these shapes into various constructors of one type, \Shape.
Future work could track the different types of shapes and the allowable transformations between them,
  which could enable more granular rewrite rules.

Mask filters, which are commonly used to apply Gaussian blurs,
  interact with shape boundaries in ways that are not yet fully captured by our model.
Future work could formalize these interactions.
This direction may be particularly relevant as
  the Skia team has acknowledged the complexity of mask filters
  and are trying to replace them with image filters to simplify the API.

In the course of exploring Skia we noticed that Skia and other 2D graphics libraries share a lot of features,
  and often have quite similar APIs too.
Hence, one direction we plan to explore in the future
  is expanding our formalization to other graphics APIs so that,
  for example, we could verify a compiler from SVG or HTML \texttt{<canvas>} to Skia or another back-end library.
We also hope to explore compositing,
 a type of caching used in animations,
 and how it impacts graphics library correctness and performance.

\section{Related Work}
\label{sec:related}

\paragraph{2D graphics}

Modern rasterization can be said to start with
  Warnock and Geschke's Interpress system at Xerox PARC~\cite{interpress}.
Warnock and Geschke then founded Adobe~\cite{about-adobe},
  which introduced PostScript~\cite{postscript}, and PDF~\cite{pdf}.
Color blending was introduced by \citet{blendmodes}.
\citet{pipeline} introduced a rudimentary rendering pipeline
  separating what is being drawn (objects and their spacial positions)
  from how it is being drawn (rendering).
The standard textbook on the topic is
  \textit{Computer Graphics: Principles and Practice},
  especially the first edition~\cite{cgpp}.
A more recent thread of development
  in the programming languages community
  is the Halide programming language~\cite{halide}.
In Halide, individual graphical kernels,
  like those that implement drawing or blending operations,
  are described as a combination of computation and \emph{schedule}.
Schedules can inline across operations, tile and vectorize operations,
  and even split operations across multiple cores or accelerators
  without affecting the computation itself.
Further work \citep{halide-auto, halide-tree}
  introduced an auto-tuner to automatically generate high-quality schedules.
Work by \citet{halide-verif},
  on automatically verifying optimizing rewrites in the Halide compiler,
  is particularly relevant to this work.
All that said, Halide and \muskia focus on different aspects of rasterization:
  Halide focuses on the implementation of individual instructions,
  while \muskia focuses on the semantics of the Skia command language itself.

\paragraph{Web Browser Optimizations}
Web browsers are prominent and demanding clients of rasterization libraries,
  and have been a focus of recent research.
Web browser JavaScript engines have seen significant work,
  dating back to  \citet{tracemonkey}'s initial JIT compiler
  for Firefox's JavaScript engine.
Layout engines are also an area of active research,
  including \citet{parlayout}'s parallel layout algorithm,
  \citet{layoutsynth}'s synthesis algorithm for web browser layout algorithms,
  and \citet{spineless}'s algorithm for optimized layout invalidation.
However, while web browsers are important clients for rasterization libraries,
  rasterization is a later phase in the browser rendering pipeline
  and does not directly interact with layout.

\paragraph{Semantics for graphical domains}

A number of papers have defined semantics for
  programming languages in other graphical domains.
\citet{conal} presents a framework for functional programming over images,
  which is strikingly similar to the abstract model presented in \Cref{subsec:abstractmodel}.
However, in this paper we go beyond this abstract model,
  and introduce a mapping of the structured semantics of a widely-used graphics API
  onto the abstract model.
\citet{cassius} formalize a significant fraction of the CSS specification,
  and later work from the same author~\cite{vizassert}
  uses that formalization to verify geometric assertions
  about web page layouts,
  such as guaranteeing that two page components do not overlap.
\citet{reincarnate} introduces denotational semantics
  for the compilation of CAD programs into 3D meshes,
  and \citet{gcode} introduce a semantics for the lower level
  G-code instructions used by 3D printing machines.
The Marshall language~\cite{marshall}
  likewise defines a semantics for CAD,
  with a focus on defining the semantics of continuous real operators.
There is also a body of work on semantics and decision procedures
  for knitting machines and knitting programs (\citet*{knitsem, knitequiv, ufo}).
While there's no direct relation with our work,
  these papers demonstrate the significant interest in graphical computing
  in the programming languages community.

\section*{Data-Availability Statement}
We will be submitting two artifacts.
The first is the Lean mechanization of
  \muskia semantics and the rewrites discussed in \Cref{sec:rewrites}.
This Lean mechanization will also include the translation validator discussed in
  \Cref{subsec:rq1}.
The second is our fork of Skia with our optimizer
  that implements the aforementioned rewrites.
We will also be providing the script we used to pull \texttt{skp} files for
  websites, along with the original 100 websites we base most of our evaluation on
  and another 100 websites used for \Cref{subsec:rq4} in this artifact.
We have made the artifacts available at \mbox{\url{https://zenodo.org/records/21386615}~\cite{artifact}}.

\section*{Acknowledgments}
We thank Thalia Archibald, Andrew Riachi, and Artem Yadrov for helpful feedback on drafts, and Artem
Yadrov additionally for testing the artifact.
We thank Brian Osman and Michael Ludwig at Google for multiple conversations clarifying Skia's
semantics, guarantees, and performance properties, and the many other Skia developers for the
quality of the Skia codebase.
We thank Yuvaraj Chesetti for preliminary work in optimizing Skia.
Finally, we thank the anonymous reviewers for suggestions that made this paper much stronger.
This work is supported by the National Science Foundation under Grant no. CCF-2340192.

\bibliographystyle{ACM-Reference-Format}
\bibliography{references}

\end{document}